\documentclass{moriond}

\def\be{\begin{equation}}
\def\ee{\end{equation}}
\def\bea{\begin{eqnarray}}
\def\eea{\end{eqnarray}}

\newcommand{\Photo}{\includegraphics[height=35mm]{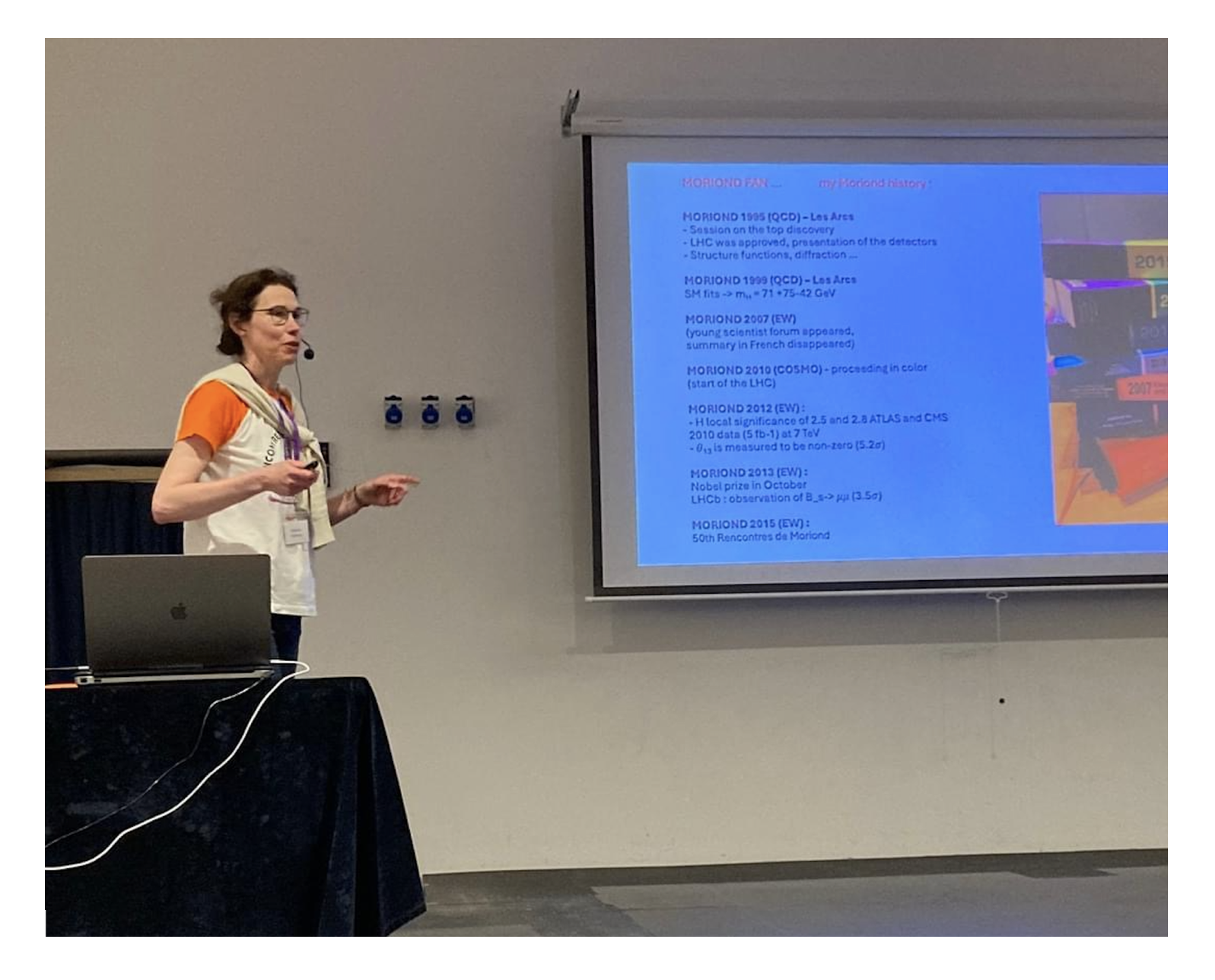}}

\begin{document}
\vspace*{4cm}
\title{Experimental Summary \\
Electroweak Interaction \& Unified Theories}

\author{ Barbara Clerbaux }

\address{IIHE - Interuniversity Institute for High Energies, ULB - Universit\'e libre de Bruxelles, \\
Boulevard du Triomphe 2, 1050 Brussels, Belgium} 

\maketitle\abstracts{
The proceeding gives a summary of the experimental results presented at the Moriond 2024 conference (electroweak interactions and unified theory session), focussing on the latest measurements. A variety of topics were covered in the various sessions of the conference including H boson properties, measurements in the electroweak and top quark sectors, flavour and neutrino physics, searches for dark matter and physics beyond the standard model, as well as astroparticle and precision physics. }

%
\section{Introduction}
\begin{figure}[h]
\begin{minipage}{0.80\linewidth}
This is my 8th participation to a Moriond conference and as always it is an exciting and inspiring moment. Surely one can say that I am part of the Moriond's fans. The conference is characterised by having plenary talks from experimentalists and theoreticians covering a variety of area and experiments, organised in different sessions. A total of 52 experimental talks has been presented during the week of the conference, each talk being followed by lively discussions. Summarising all these presentations is clearly a challenge, and in this proceeding the focus will be given on the new and latest results of the various sessions. Note that in addition 20 short presentations of high quality were given in the three "young scientist forum" sessions. All the slides of the talks are available on the web site of the conference.
\end{minipage}
\begin{minipage}{0.17\linewidth}
\centerline{\includegraphics[width=0.9\linewidth]{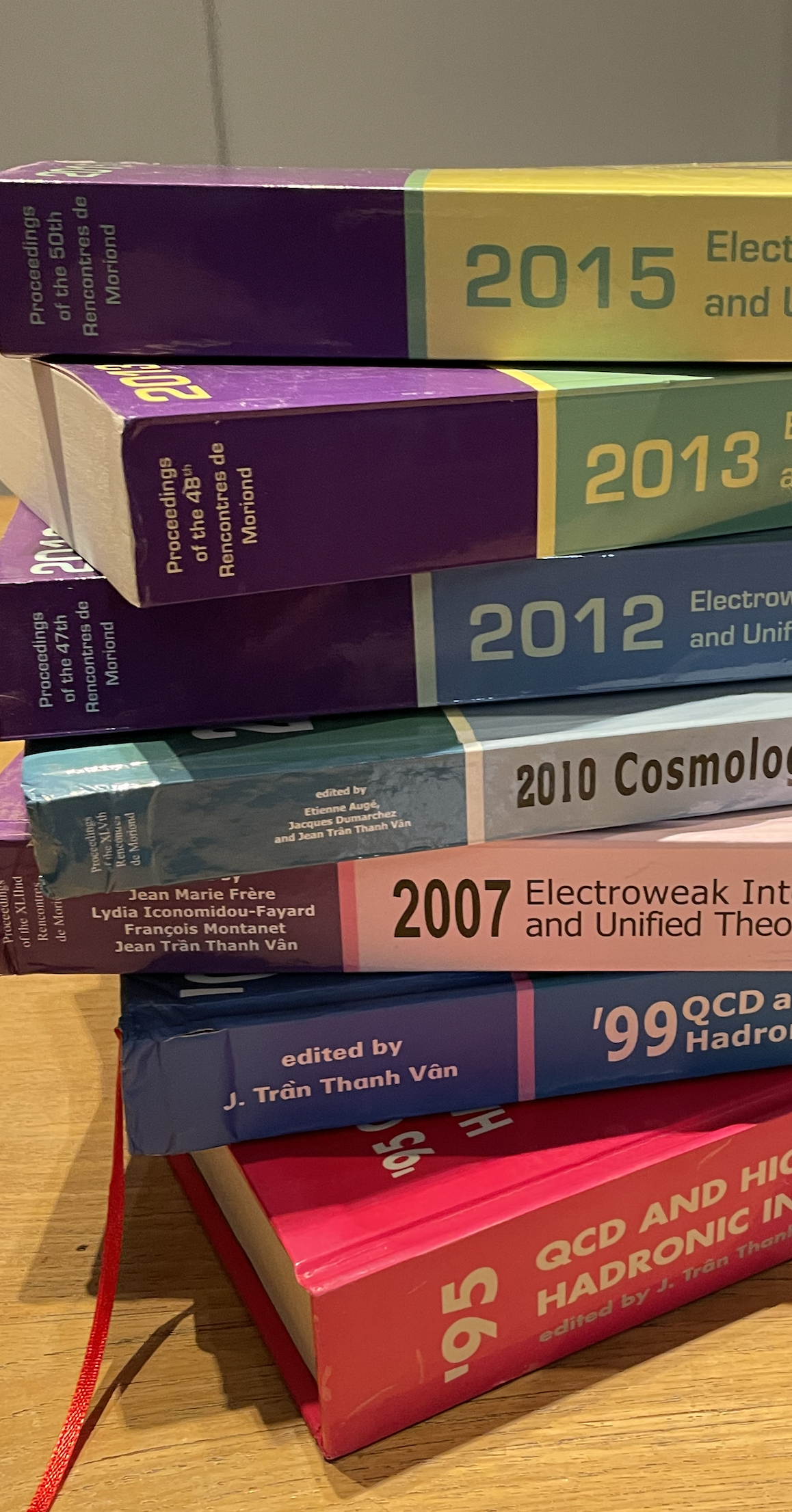}}
\end{minipage}
\end{figure}
The proceeding is organised as follows. After this short introduction, the second section details results on the H boson sector.
The third and fourth sections are dedicated to precise standard model (SM) measurements in the electroweak sector and the top quark sector, respectively. Section 5 gives a short summary of searches beyond the standard model (BSM), followed by the results on flavour physics in Section 6. Finally the next two sections present the latest results on neutrino physics (Section 7) and dark matter searches and precision measurements (Section 8). The last section presents the conclusions.
%
\section{H boson results} \label{sec:Hboson}

The discovery of the H boson at CERN in 2012 by the ATLAS and CMS experiments was a key achievement of the community. The H boson was the last elementary particle of the SM discovered, confirming the Brout-Englert-Higgs mechanism of the electroweak symmetry breaking proposed 48 years earlier. The nature and the properties of the H boson is up to now in agreement within uncertainties with the ones predicted by the SM, however room for new physics is still present in many area and more precise measurements are still to come. In this section, we give first a short status of the LHC data taking and future programs, as the H boson sector is studied mainly by the ATLAS and CMS experiments.  A selection of results from the H boson session is then presented, in particular in what concerns its mass and width measurements, the differential cross section measurements, the di-H production and the search for new physics in the H sector.

\vspace{0.1cm}
\textbf{The Large Hadron Collider (LHC)} at CERN took proton-proton interaction data in several periods at different centre of mass energies: the run 1 (data taken in 2010-2012, at 7 and 8 TeV, recorded integrated luminosity of about 25 fb$^{-1}$), the run 2 (data taken in 2015-2018, at 13 TeV, luminosity of about 140 fb$^{-1}$) and run 3 (data taking period 2022-25 at 13.6 TeV). Up to now about 135 fb$^{-1}$ have been recorded by ATLAS and CMS for the run 3, leading to date to an impressive total luminosity of 300 fb$^{-1}$ per experiment for the three runs. More data will come by the end of the run 3. The high luminosity LHC (HL-LHC, also called LHC phase 2) data taking will take place from 2029 to 2041 (runs 4-6) with an expected total luminosity of about 3000 fb$^{-1}$ and with major upgraded detectors for ATLAS and CMS. The main physics program of the HL-LHC is to study the H sector in much more precision.

\begin{figure}[b!]
\begin{minipage}{1.0\linewidth}
\centerline{\includegraphics[width=0.55\linewidth]{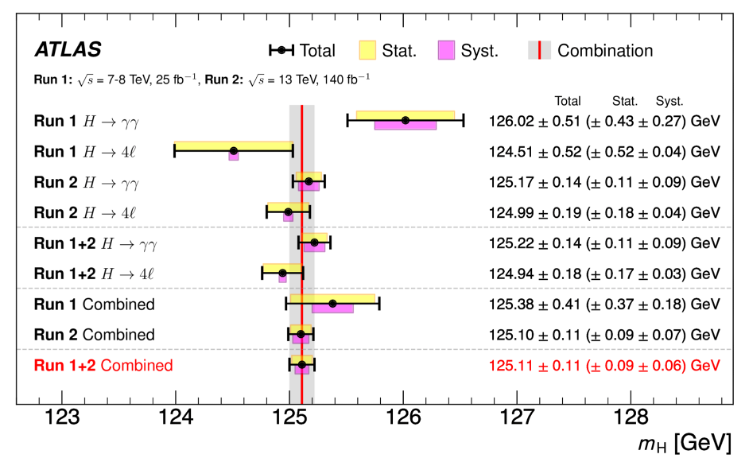}}
\end{minipage}
\caption[]{The ATLAS H mass measurements for various decay channels using run 1 and run 2 datasets and their combination.}
\label{fig:Hmass_ATLAS}
\end{figure}

\vspace{0.1cm}
\textbf{The H mass and width} are fundamental parameters of the SM. The latest (full run 2) mass measurement from CMS and from ATLAS are $m_{\rm H}=125.04 \pm 0.11$(stat)$\pm0.05$(syst) GeV for the H$\rightarrow$ZZ$\rightarrow$4$\ell$ decay channel and 
$m_{\rm H}=125.17 \pm 0.11$(stat)$\pm0.09$(syst) GeV for the H$\rightarrow$$\gamma \gamma$ decay channel, respectively~\cite{Devivie}, 
the main uncertainties coming from the lepton and photon energy scales.
Figure~\ref{fig:Hmass_ATLAS} presents the various H mass measurements of ATLAS and the final run 1 and run 2 combination, leading to a relative precision of 0.09\%. The H mass uncertainty target for the HL-LHC is about 20 MeV.
The tiny width predicted in the SM of 4.1 MeV is much smaller than the experimental mass resolution of about 1 to 2 GeV. However BSM contributions could bring a significant enhancement of the H width. ATLAS and CMS deduced an indirect limit on the H width using the ratio 
of the off-shell and on-shell cross section measurements.

\vspace{0.1cm}
\textbf{The H boson couplings} to the most massive particles are well established and evidence for the H$\rightarrow$ $\mu \mu$ decay channel was presented in 2020. The next challenge for ATLAS and CMS is the measurement of the  H to charm quark pair coupling. CMS provided a first observation of VZ with Z$\rightarrow$c$\bar{\rm c}$  production at a hadron collider at 5.7$\sigma$ and an upper limit at 95 \% confidence level (CL) on the product of the VH cross section and the branching ratio BR(H$\rightarrow$c$\bar{\rm c}$) at 14 times the SM value~\cite{Trevisani}. The ttH production with H$\rightarrow$b$\bar{\rm b}$ allows to measure both the top and the bottom coupling values. The ATLAS and CMS results show a slight deficit of events (however with large errors) compared to the SM prediction: the signal strength, defined as the ratio of the cross section to its SM prediction, $\mu$= $\sigma/\sigma_{\rm SM}$ is measured to be 0.35$\pm$0.35 for ATLAS and 0.33$\pm$0.26 for CMS. New observed (expected) upper limits of 3.7 (6.1) times the SM at 95\% CL was obtained by CMS for the bbH challenging process, with H$\rightarrow$WW and H$\rightarrow$$\tau \tau$. 

\begin{figure}
\begin{minipage}{0.45\linewidth}
\centerline{\includegraphics[width=0.99\linewidth]{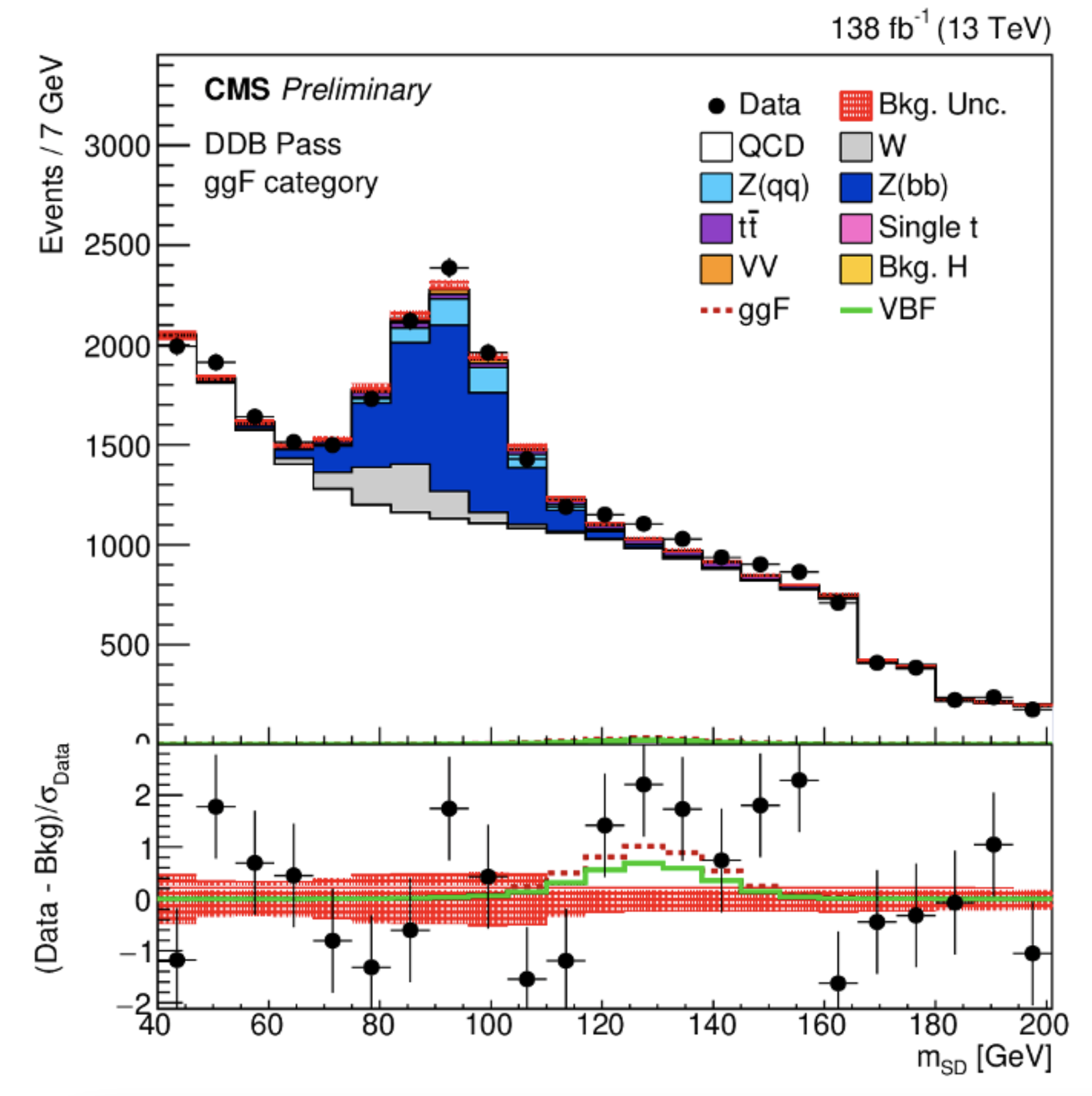}}
\end{minipage}
\hfill
\begin{minipage}{0.50\linewidth}
\centerline{\includegraphics[angle=0,width=0.99\linewidth]{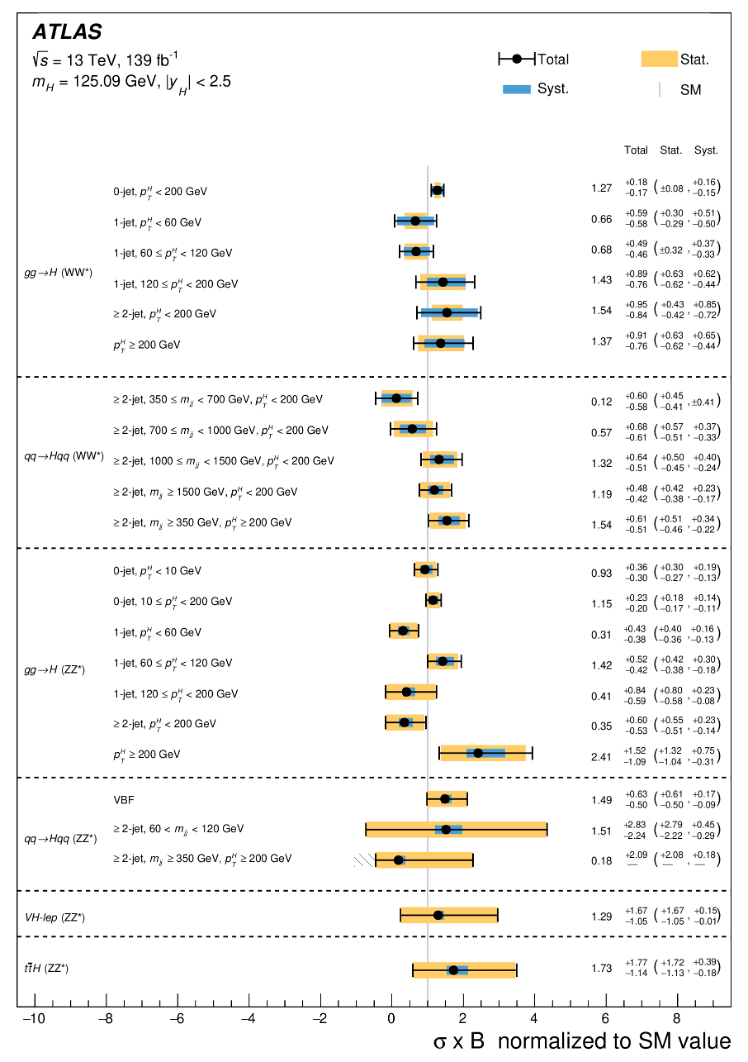}}
\end{minipage}
\caption[]{The b quark pair mass spectrum of the new CMS H$\rightarrow$b$\bar{\rm b}$  analysis (left). The analysis is performed in the boosted b$\bar{\rm b}$ regime ($p_{\rm T}$(H)$>$450 GeV), using a dedicated double-b-quark tagger called DeepDoubleB; ATLAS H cross section measurements normalised to its SM value for various $p_{\rm T}$(H) and $N_{\rm jet}$ bins (right).}
\label{fig:Hcrosssection}
\end{figure}

\vspace{0.1cm}
\textbf{Fiducial differential cross section measurements}  were presented for various decay channels (H$\rightarrow$$\gamma \gamma$, H$\rightarrow$ZZ,WW,  H$\rightarrow$b$\bar{\rm b}$ and H$\rightarrow$$\tau \tau$)~\cite{Mohammadi}. Indeed the large data sample collected during the run 2 allows to go differential. The cross sections are measured as a function of the number of jets $N_{\rm jet}$ (sensitive to the production mode composition) and the transverse momentum $p_{\rm T}$ (sensitive to QCD perturbative modelling and to BSM scenarios). As an example, the new CMS H$\rightarrow$b$\bar{\rm b}$ analysis (in the boosted regime) is presented in Figure~\ref{fig:Hcrosssection}(left) and the ATLAS signal strengths of a variety of H processes in  Figure~\ref{fig:Hcrosssection}(right).

\begin{figure}
\begin{minipage}{0.50\linewidth}
\centerline{\includegraphics[width=0.9\linewidth]{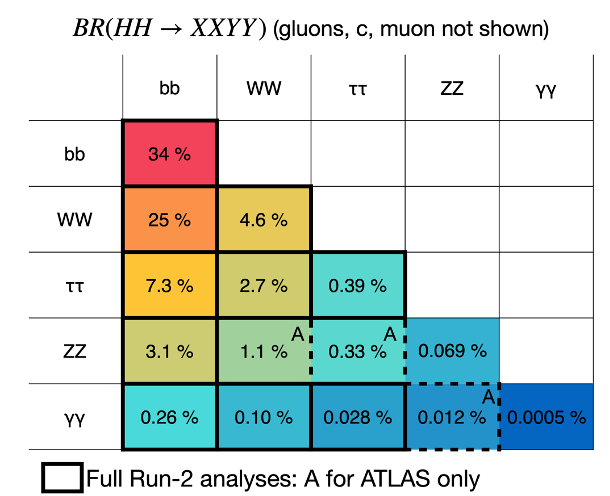}}
\end{minipage}
\hfill
\begin{minipage}{0.50\linewidth}
\centerline{\includegraphics[angle=0,width=0.9\linewidth]{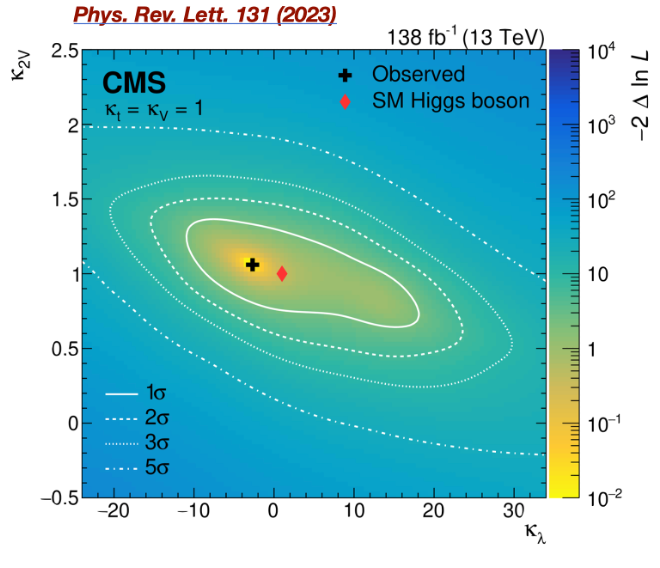}}
\end{minipage}
\caption[]{The various final states and branching ratio for HH$\rightarrow$XXYY decays; Results of the CMS HH$\rightarrow$b$\bar{\rm b}$b$\bar{\rm b}$ analysis (left).}
\label{fig:HH}
\end{figure}

\vspace{0.1cm}
\textbf{The HH processes} are particularly interesting to measure as they are directly connected to probing the H potential and in particular the trilinear coupling $\lambda_{\rm HHH}$. The HH invariant mass shape is very dependent on the coupling constant ratio $\kappa_{\lambda}$=$\lambda_{\rm HHH}$/$\lambda_{\rm SM}$.
The HH production cross section is typically 1000 time smaller than the one for single H production, which makes it an extremely rare process at the LHC.
Depending on the H decays, an impressive list of final states was covered by ATLAS and CMS, as shown in Figure~\ref{fig:HH}(left)~\cite{Deramo}. Thanks to sophisticated analysis tools the HH searches for the leading 3 decay channels (b$\bar{\rm b}\gamma \gamma$, b$\bar{\rm b}\tau \tau$, b$\bar{\rm b}$b$\bar{\rm b}$) have each an expected limit as low as around 5 x SM prediction, and about 2.5-3 x SM when combined. New results were presented by ATLAS and CMS for the VBF HH$\rightarrow$b$\bar{\rm b}$b$\bar{\rm b}$ and HH$\rightarrow$$\gamma \gamma$$\tau \tau$ analyses, respectively. The HH vector boson fusion (VBF) production mode is sensitive to the VVHH couplings $\kappa_{\rm 2V}$. The CMS HH$\rightarrow$b$\bar{\rm b}$b$\bar{\rm b}$ boosted analysis excludes the value $\kappa_{\rm 2V}$=0 with a significance of 6.3$\sigma$ as shown in Figure~\ref{fig:HH}(right). The HH analyses are also sensitive to potential presence of new physics, in particular to new spin 0 or spin 2 particles.

\vspace{0.1cm}
\textbf{Search for new physics in the H boson sector} is an active area for ATLAS and CMS. A wide scope of new BSM H boson searches has been released by ATLAS and CMS. No excess are observed above the SM prediction, however still a large amount of phase space is available for extended H sectors~\cite{Uttley}. In the search for low mass H$\rightarrow$$\gamma \gamma$, CMS observes an excess of local (global) significance of 2.9$\sigma$ (1.3$\sigma$) at a mass of 95.4 GeV, ATLAS observes a local significance of 1.7$\sigma$ at 95.4 GeV. 

\section{Electroweak precision measurements} \label{sec:EWpres}
Precise measurements in the electroweak sector are key to test the SM. Although the LHC is a hadron machine, precise electroweak measurements have been obtained by ATLAS, CMS and LHCb. The latest measurements of the W and Z cross sections and the W mass are presented, followed by two beautiful wildcards by ATLAS and CMS, and electroweak results from two-photon collisions. 

\begin{figure}
\begin{minipage}{0.45\linewidth}
\centerline{\includegraphics[width=0.99\linewidth]{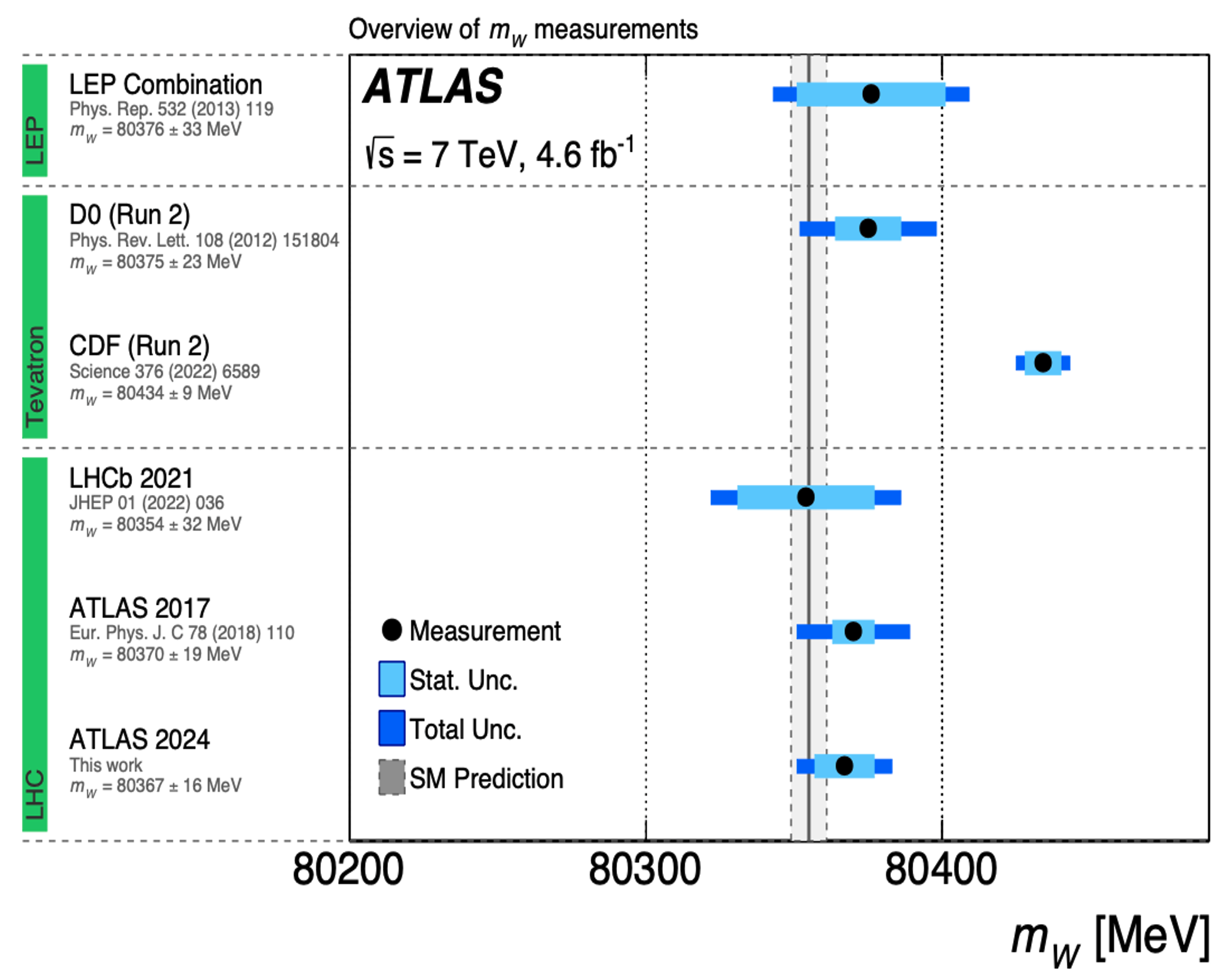}}
\end{minipage}
\hfill
\begin{minipage}{0.53\linewidth}
\centerline{\includegraphics[angle=0,width=0.99\linewidth]{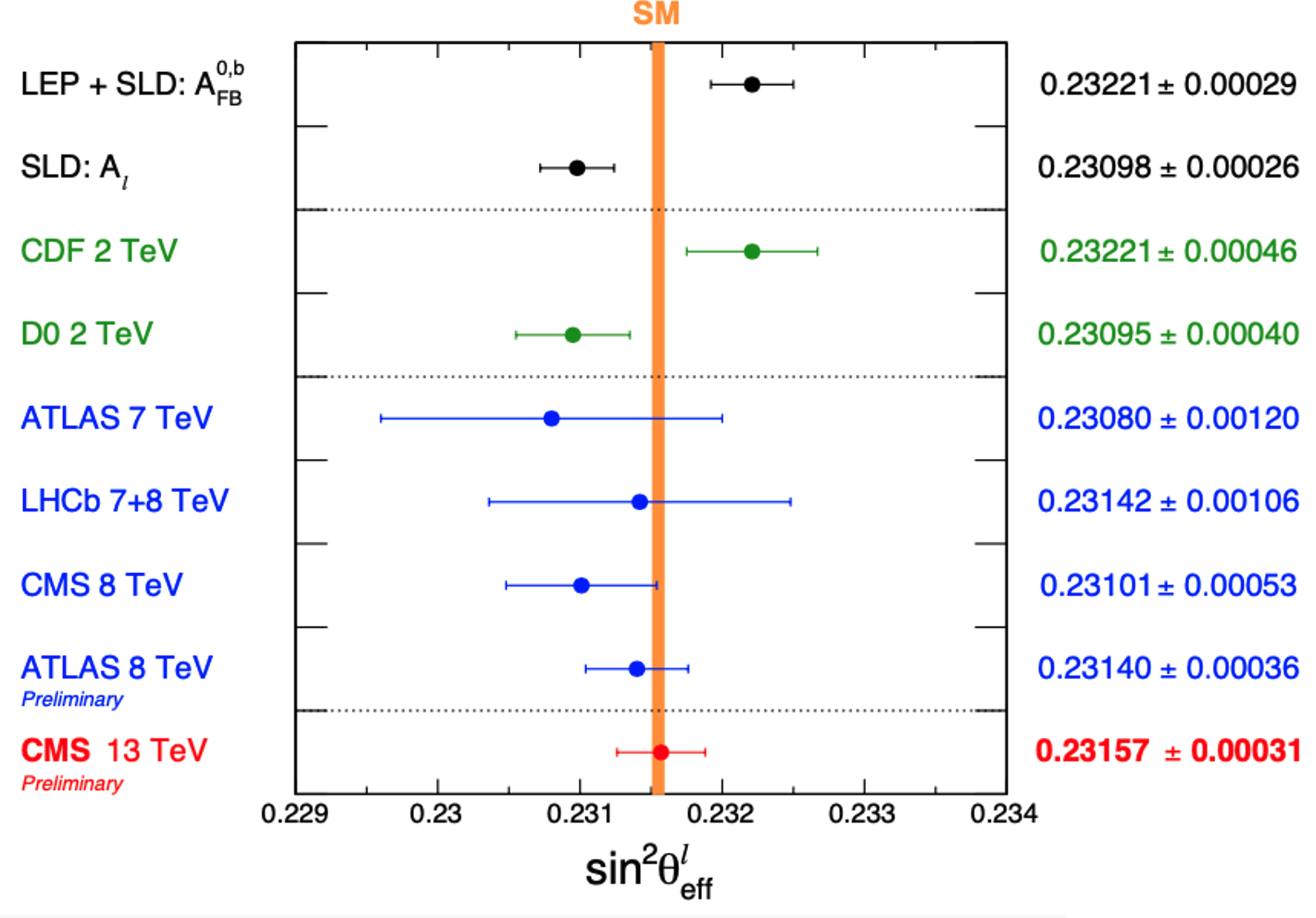}}
\end{minipage}
\caption[]{Latest W mass measurement by ATLAS, together with previous results (left); CMS  measurement of $\sin^2\theta^\ell_{\rm eff}$ together with older results (right).}
\label{fig:EW1}
\end{figure}

\vspace{0.1cm}
\textbf{Various W and Z boson measurements} from ATLAS, CMS and LHCb were presented. The very fresh results of Z and W cross section production at 13.6 TeV from the run 3 partial dataset analysis show a good agreement with the SM. ATLAS updated its previous analysis of the W mass with extended studies of the PDF (parton density functions) to reduce as much as possible the uncertainties~\cite{Long}. The W mass is extracted from the W boson transverse mass and $p_{\rm T}$ distributions. The obtained value, $m_{\rm W} = 80366.5 \pm 15.9$ MeV, has an impressive precision of less than 0.02\%. The W mass result, shown in Figure~\ref{fig:EW1}(left) is in good agreement with the SM and does not confirm the higher value of the W mass obtained in 2022 by the CDF data re-analysis measurement. The ATLAS analysis is also sensitive to the W width, measured to be $\Gamma_{\rm W} = 2202 \pm 47$ MeV.
ATLAS performed a comprehensive study of events with jets and large missing transverse energy (MET) in the final state, providing a measurement of the differential 
Z$\rightarrow \nu \bar{\nu}$ cross section as a function of the  Z boson $p_{\rm T}$. The W mass is also measured by LHCb with uncertainties that are anti-correlated to that of ATLAS and CMS. Using about a third of the available run 2 dataset, the value of $m_{\rm W}  = 80354 \pm 32$ MeV is obtained by LHCb, with the target to have an expected statistical precision with the full run 2 dataset of about 14 MeV~\cite{Merli}.

\vspace{0.1cm}
\textbf{CMS Wildcard - The $\sin^2\theta^\ell_{\rm eff}$ measurement} is a CMS highlighted new result presented at the conference. The mixing angle $\sin^2\theta^\ell_{\rm eff}$ is a key parameter of the SM and is calculated using other precise experimental inputs to be $\sin^2\theta^\ell_{\rm eff}$(SM)=0.23155$\pm$0.00004. Up to now the most precise measurements come from LEP and SLD, and differ between each other by about 3$\sigma$. The new CMS analysis uses the Drell-Yan events with electron or muon pairs in the final state~\cite{Khukhunaishvili}. In case of the electron channel, the very forward calorimeters up to a pseudorapidity value of $|\eta|= 4.36$ are added in the event selection, increasing significantly the measurement precision of the forward-backward asymmetry in the lepton decay angle. From this, a value of $\sin^2\theta^\ell_{\rm eff}$=0.23157$\pm$0.00031 is extracted, reaching a comparable precision as the LEP and SLD measurements, as shown in Figure~\ref{fig:EW1}(right).

\begin{figure}
\begin{minipage}{0.48\linewidth}
\centerline{\includegraphics[width=0.99\linewidth]{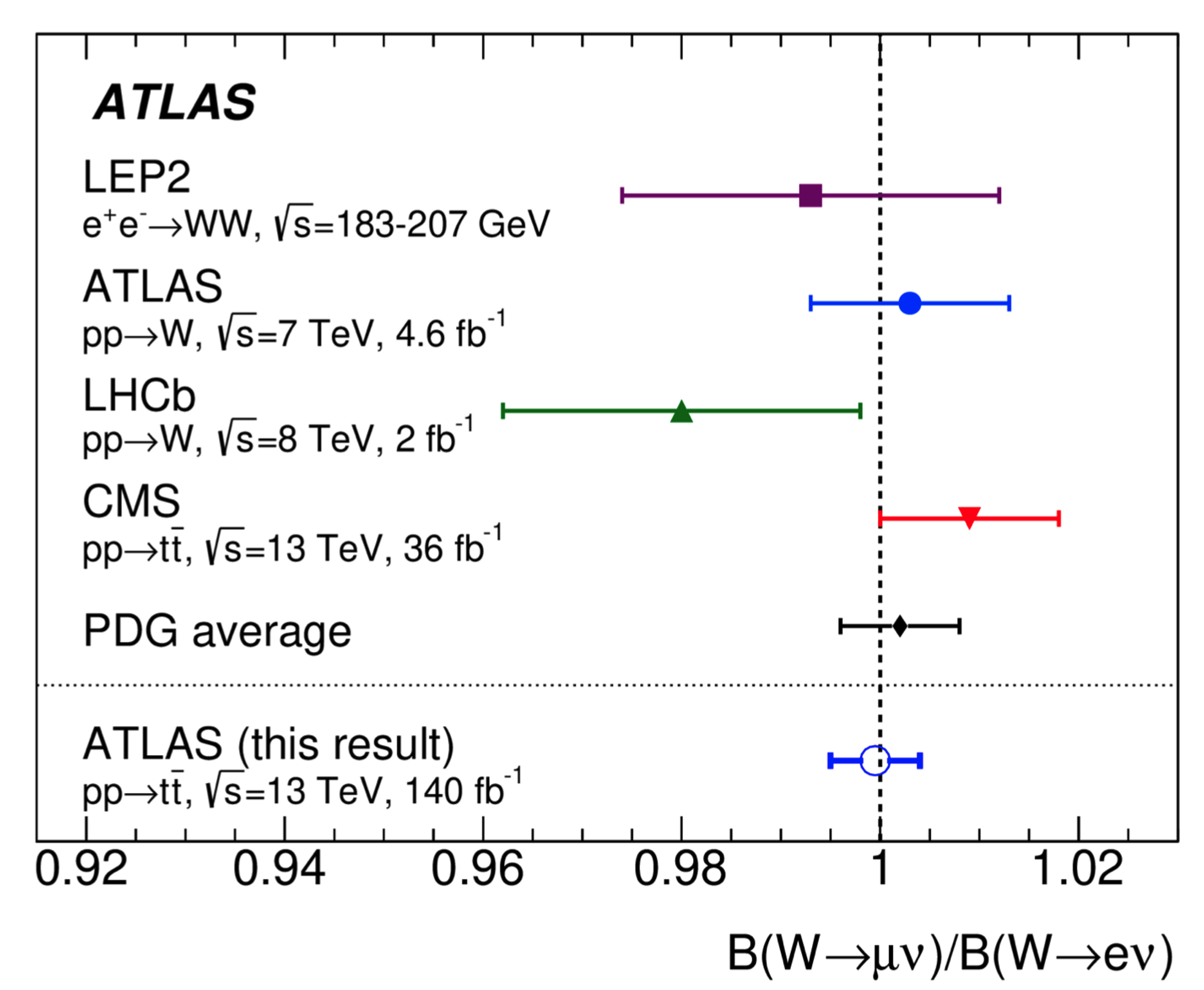}}
\end{minipage}
\hfill
\begin{minipage}{0.48\linewidth}
\centerline{\includegraphics[angle=0,width=0.99\linewidth]{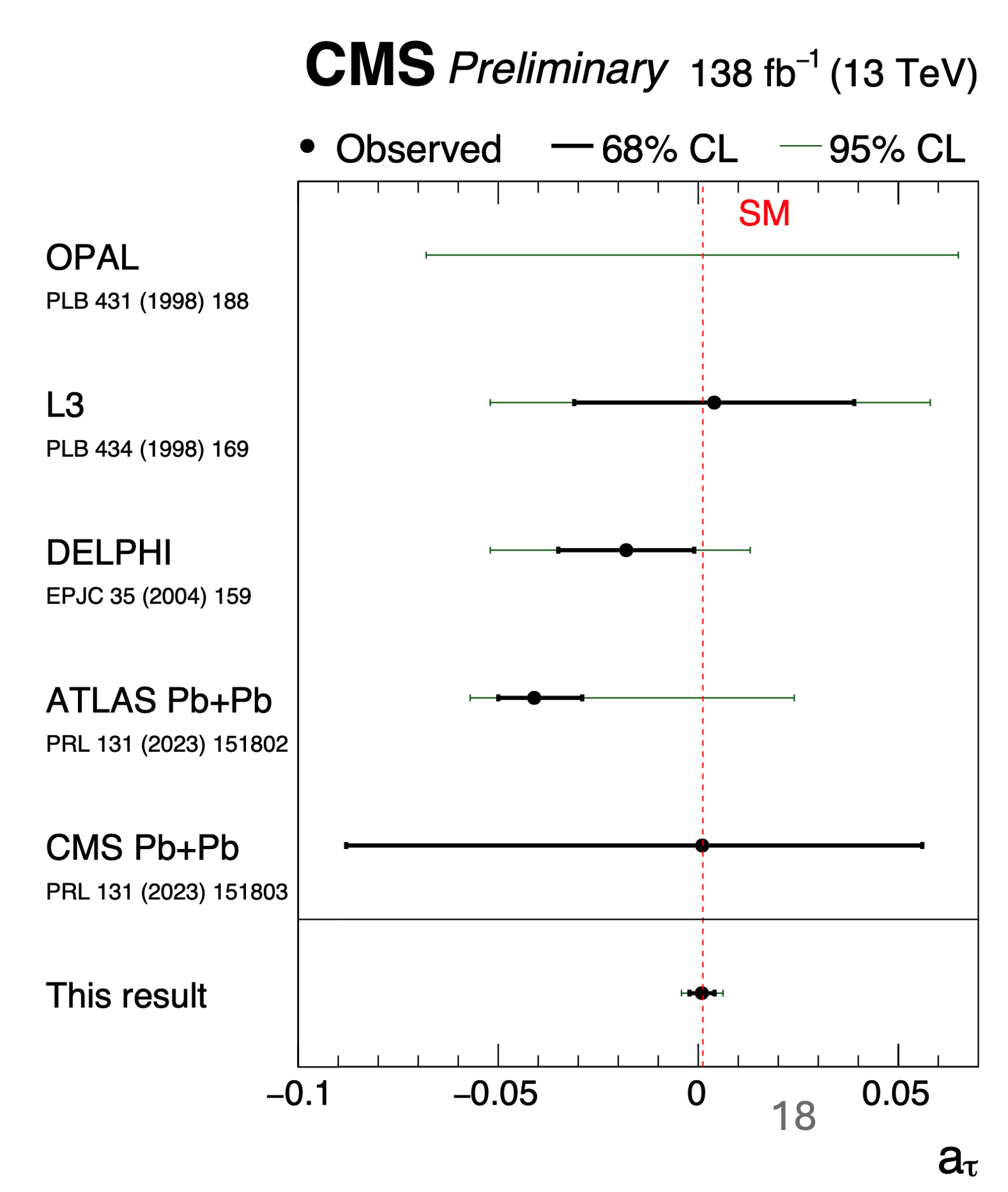}}
\end{minipage}
\caption[]{ATLAS test of the lepton flavour universality via the measurement of the ratio R = BR(W$\rightarrow$$\mu\nu$ )/ BR(W$\rightarrow$e$\nu$) together with previous results (left); Measurement of the tau lepton anomalous magnetic moment a$_\tau$ by CMS, compared with previous results (right).}
\label{fig:EW2}
\end{figure}

\vspace{0.1cm}
\textbf{The ATLAS wildcard - The test of lepton flavour universality (LFU) in W decays} is the highlighted new result by ATLAS~\cite{Knue}. The analysis uses the top-quark pair events and compares the occurrence of W decays in the muon and the electron final states. To reduce as much as possible the systematics uncertainties, the ratio R = BR(W$\rightarrow$$\mu\nu$)/ BR(W$\rightarrow$e$\nu$) is measured and normalised to the corresponding ratio for the Z boson BR(Z$\rightarrow$$\mu\mu$)/ BR(Z$\rightarrow$ee). The ratio R obtained is presented in Figure~\ref{fig:EW2}(left), the value is in agreement with 1 with a relative uncertainty of 0.45\%. This is the most precise single measurement for this ratio to date and is also more precise than the previous PDG (particle data group) average.

\vspace{0.1cm}
\textbf{The photon-induced production of a pair of tau leptons} is observed for the first time in proton-proton collisions by CMS at 5.3$\sigma$~\cite{Caillol}. Photon-induced processes are isolated by selecting events with a small number of charged hadrons associated with the di-tau vertex. 
Modifying the tau lepton magnetic moment modifies the $\gamma\gamma \rightarrow \tau \tau$ cross section and modifies the $p_{\rm T}$ and mass distributions of the signal. A very precise measurement of the tau lepton anomalous magnetic moment is extracted and presented in Figure~\ref{fig:EW2}(right), in good agreement with the expected SM value given as the dashed vertical line. The measurement does not show evidence for the presence of new physics that would modify its value. 
%

\section{Top quark physics} \label{sec:top}

The top quark is the heaviest elementary particle in nature, with a Yukawa coupling close to 1. The top quark is particularly interesting to study as it decays before forming an hadron, which allows physicists to access the bare quark properties. It is an excellent setting for testing perturbative QCD predictions and a potential portal to new physics. This section presents various measurements of the top quark properties (top quark mass and cross sections) and is followed by the new results on quantum entanglement in top events.

\vspace{0.1cm}
\textbf{Measurements of top quark properties} have been reported by ATLAS and CMS~\cite{Wuchterl}. Both experiments measured the inclusive top-antitop (t$\bar{\rm t}$) production cross section with the run 3 partial dataset at 13.6 TeV. A new combination of the ATLAS and CMS top quark mass measurements leads to $m_{\rm t}=172.52\pm0.14$(stat)$\pm0.30$(syst) GeV, the dominant systematics uncertainty coming from the b-quark jet energy scale. New results on top associated production were also  presented~\cite{Boumediene}. Using a deep neural network (DNN) to classify the events in three categories: ttZ+tWZ, tZq and the backgrounds, CMS presented a simultaneous measurement of the ttZ+tWZ cross section, as a function of the tZq cross section. The ttZ+tWZ cross section measurement has a small tension with the SM prediction (being slightly above at a $2\sigma$ level). The new ATLAS result on the t$\bar{\rm t}\gamma$ production is in agreement with the SM.
\begin{figure}[t!]
\begin{minipage}{0.53\linewidth}
\centerline{\includegraphics[width=0.99\linewidth]{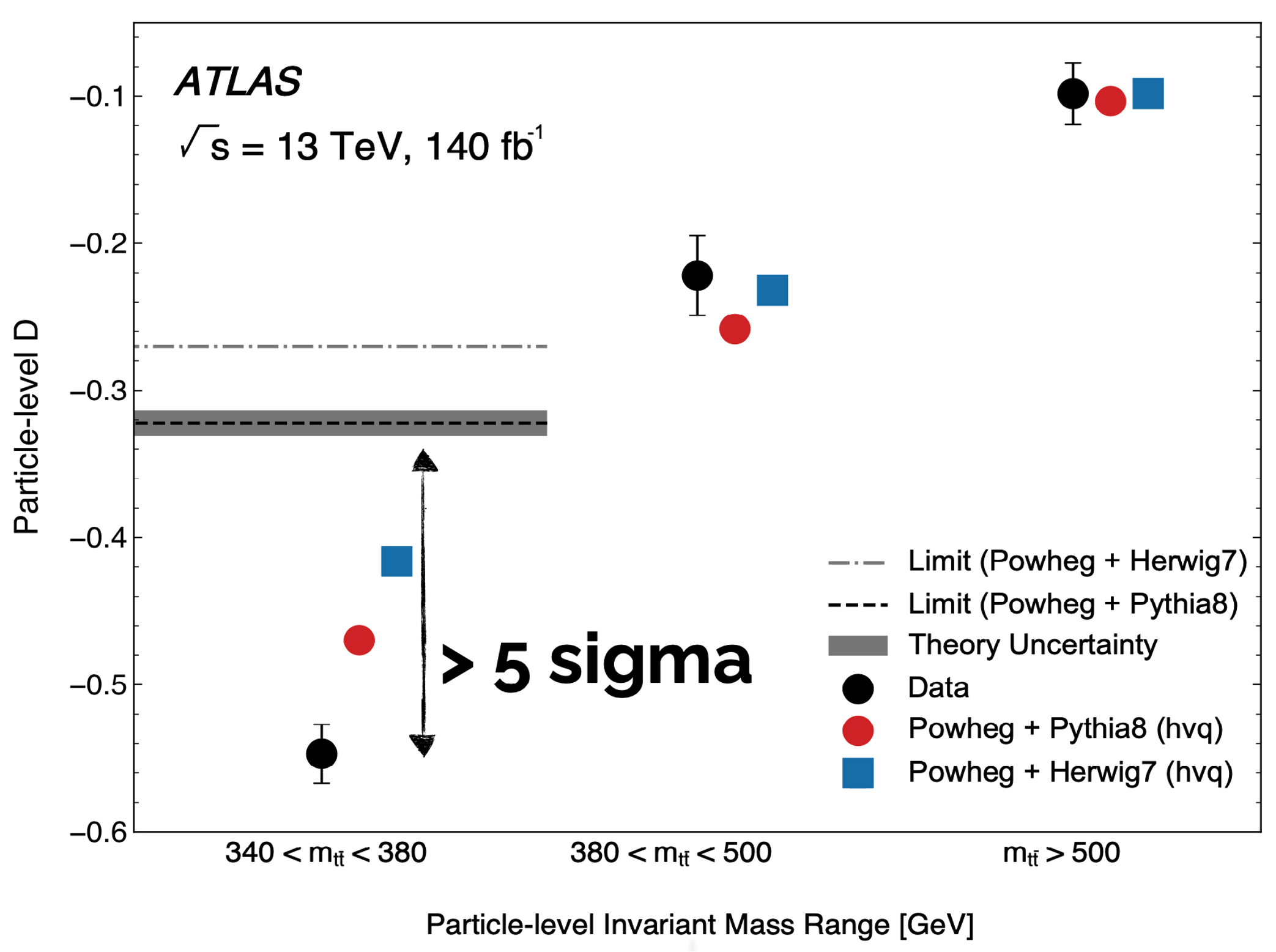}}
\end{minipage}
\hfill
\begin{minipage}{0.45\linewidth}
\centerline{\includegraphics[angle=0,width=0.9\linewidth]{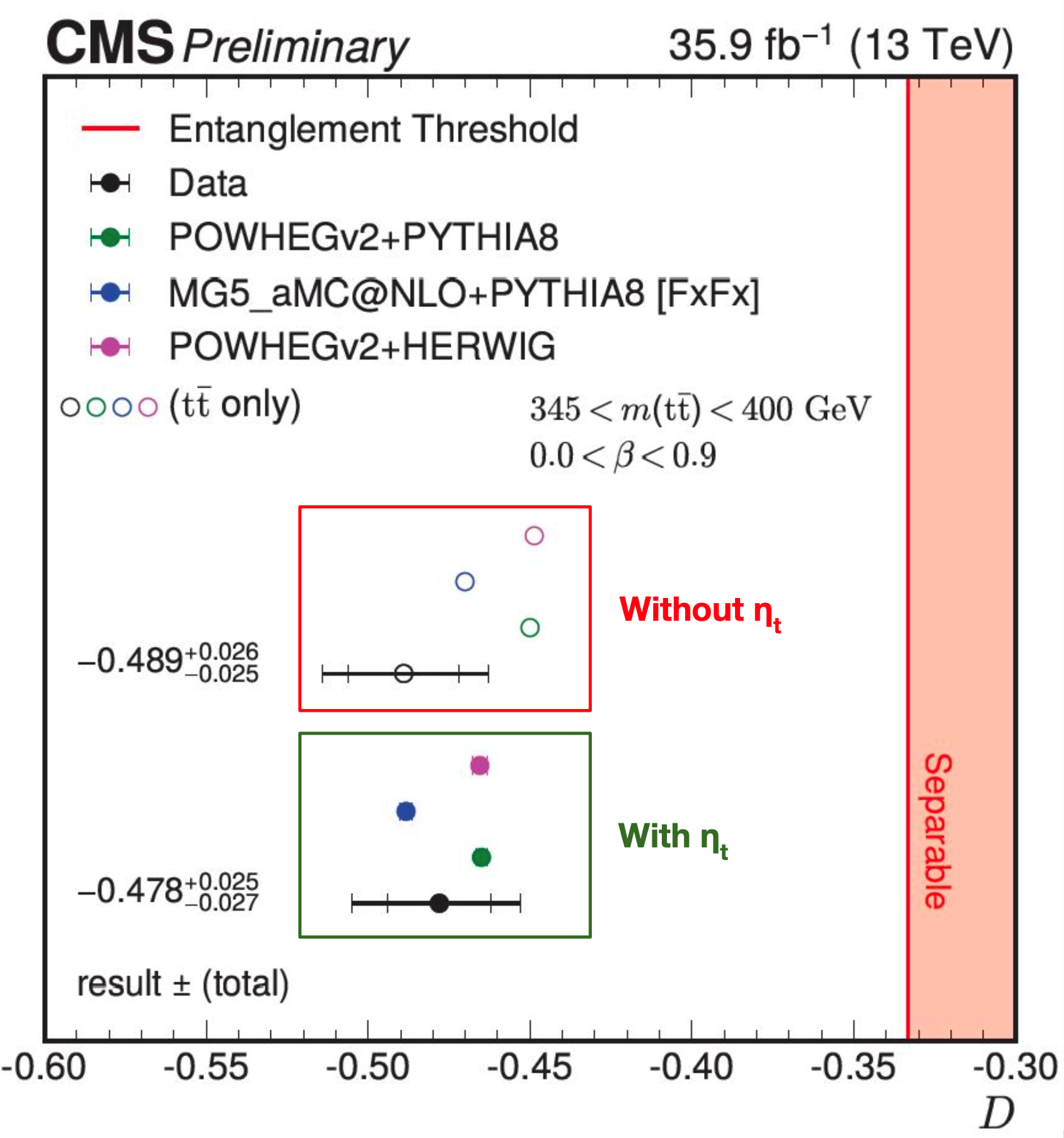}}
\end{minipage}
\caption[]{Measurements of the top entanglement, via the $D$ variable by ATLAS (left) and CMS (right). A system is considered as being in a quantum entanglement state if $D < - 1/3$. For CMS, measurements with and without the addition of a toponium state $(\eta_{\rm t})$ are presented.}
\label{fig:top}
\end{figure}

\vspace{0.1cm}
\textbf{Quantum entanglement in top events} are new results that generated excitement and discussion during the conference.
ATLAS and CMS presented their latest analysis results~\cite{Wuchterl,Bragagnolo} from top-antitop events in the dilepton decay channels. Top-quark pairs at the LHC are mainly unpolarised, with their spins being strongly correlated. The spin information can be measured via the final state particle angular variables. The spin correlation depends on the mass of the top-antitop system $m_{\rm t\bar{t}}$ and on the angular variables. A system is considered as being in a quantum entanglement state if $D < - 1/3$, where $D$ is defined as the trace of the spin correlation matrix divided by 3. The $D$ parameter can be extracted from the $\phi$ angular distribution with the relation $D=-3  \cos\phi$, where  $\phi$  is the angle between the 2 final state leptons in the top-antitop rest frame. The ATLAS and CMS results are presented in Figure~\ref{fig:top}, showing that entanglement is observed with $>5\sigma$ at low $m_{\rm t\bar{t}}$. The CMS analysis shows in addition that when a t$\bar{\rm t}$ bound state (toponium, a colour singlet pseudo-scalar state) is included in the simulation, the agreement between the measurement and the SM simulations improves in the threshold mass region.

\section{Beyond the Stantard Model} \label{sec:BSM}

Search for BSM physics is one of the key activities of particle physicist today. Indeed  despite its incredible success, the SM is clearly not a complete theory, as it does not provide a framework to describe several important observations in the universe as the presence of dark matter, the matter-antimatter asymmetry and the non-zero masses of  the neutrinos established by the neutrino oscillation measurements. Direct searches for BSM physics remain an important part of the LHC physics program. The searches cover a wide range of experimental signatures and model interpretations. Key BSM activities concern the searches for heavy resonances, leptoquarks, vector-like quarks (VLQ), heavy neutral leptons (HNL), long-lived particles (LLP), and heavy scalars~\cite{Grancagnolo,Santanasasio,Kay}. Up to now, no deviation from the SM expectation has been observed and in the following, three specific BSM activities are summarised: the ATLAS wildcard on search for spin-0 resonances, an overview on the effective field theory (EFT) and new idea of model-agnostic searches for new physics.

\begin{figure}[t!]
\begin{minipage}{0.53\linewidth}
\centerline{\includegraphics[width=0.99\linewidth]{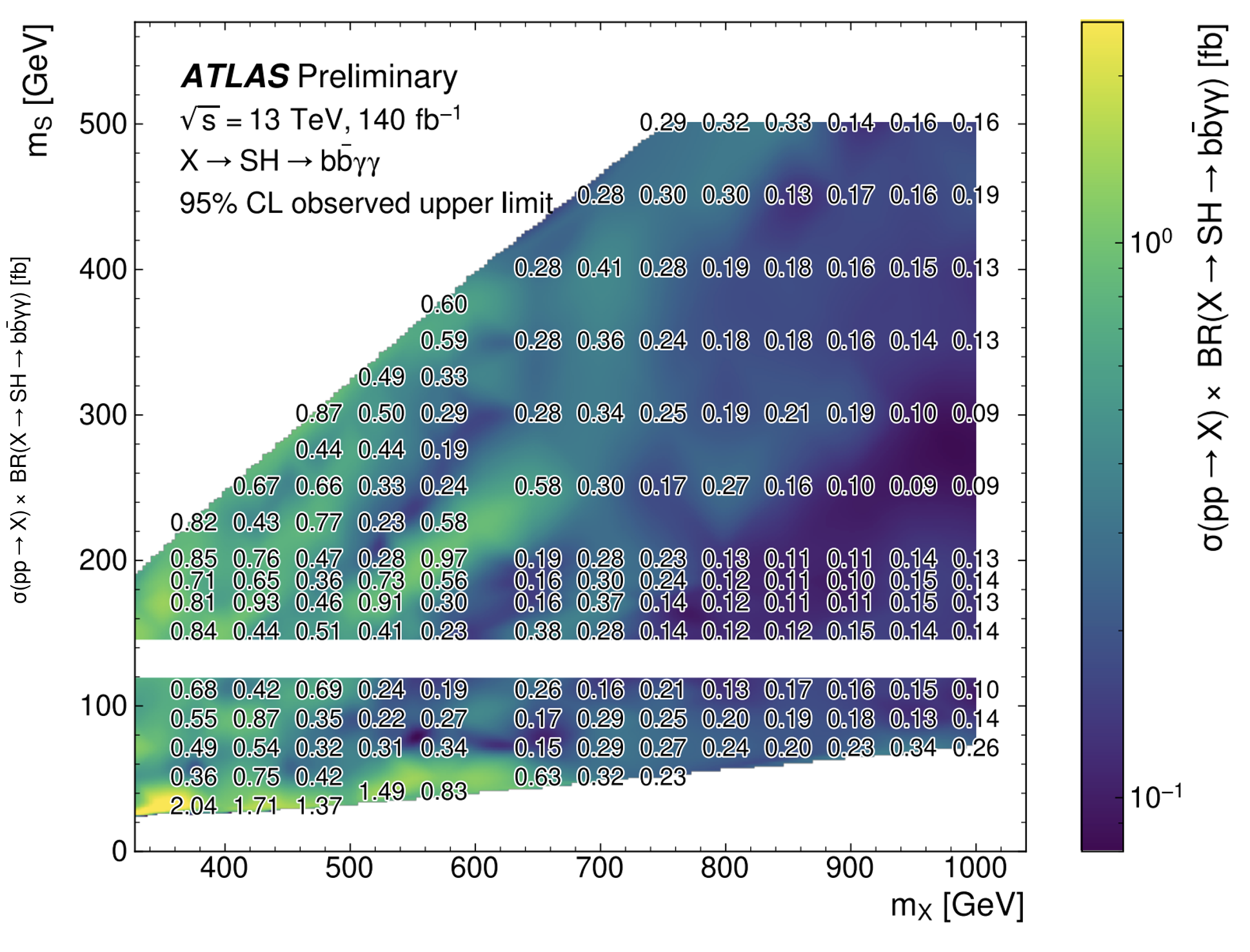}}
\end{minipage}
\hfill
\begin{minipage}{0.42\linewidth}
\centerline{\includegraphics[angle=0,width=0.99\linewidth]{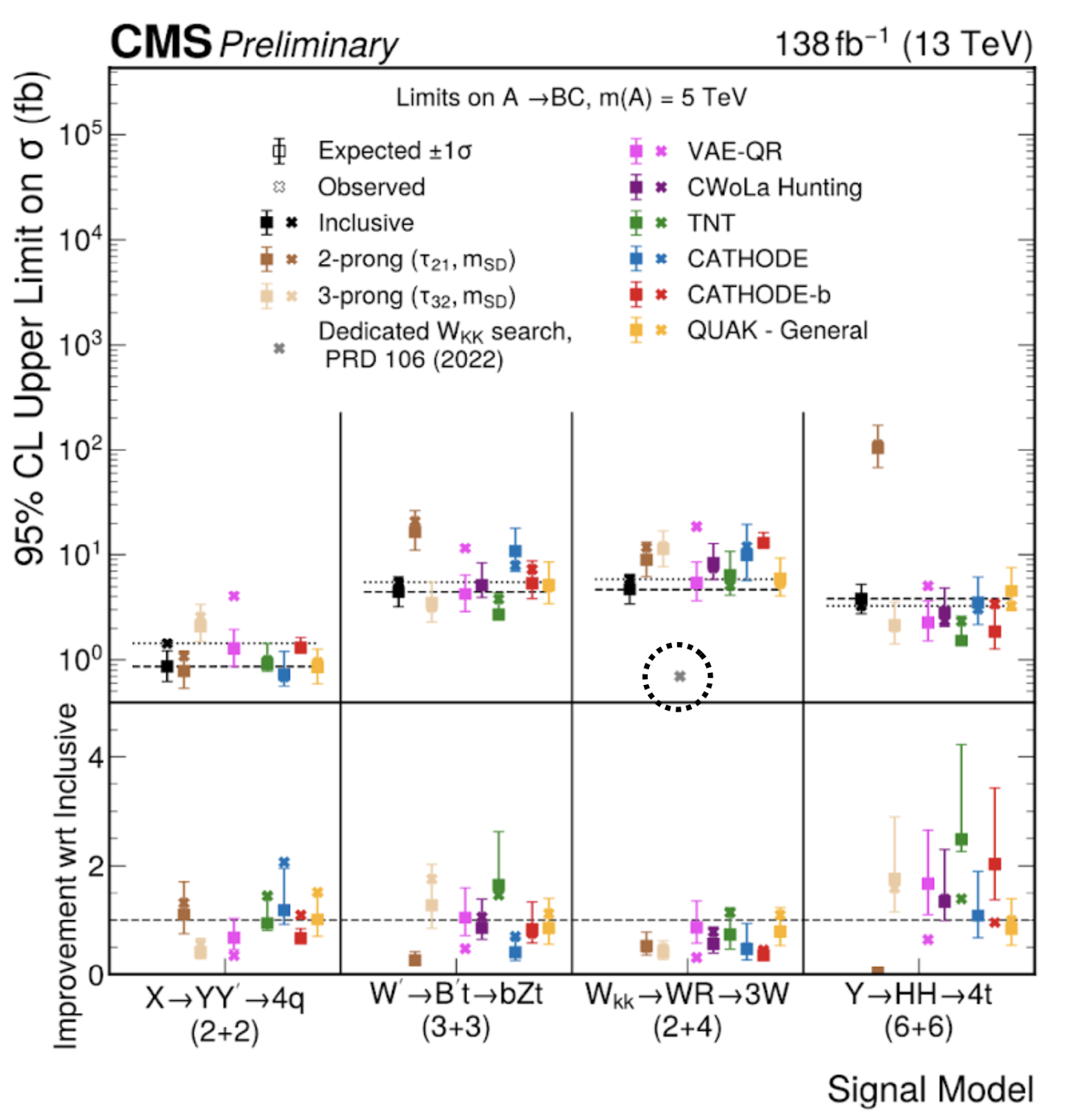}}
\end{minipage}
\caption[]{ATLAS upper limits at 95\% CL on the product of the cross section and the branching ratio $\sigma$(pp$\rightarrow$X)BR(X$\rightarrow$SH$\rightarrow$b$\bar{\rm b}\gamma\gamma$) (left); CMS upper limits at 95\% CL on the cross section for the process A$\rightarrow$BC for several anomaly detection search methods (colored points) applied to a variety of signal models (the columns), for a resonance mass $m$(A)=5 TeV.}
\label{fig:BSM}
\end{figure}

\vspace{0.1cm}
\textbf{ATLAS Wilcard - A search for new spin-0 resonances} was reported by ATLAS in the decay channel X$\rightarrow$SH$\rightarrow$b$\bar{\rm b}\gamma\gamma$~\cite{Clement}. Two signal regions targeting resolved and boosted S$\rightarrow$b$\bar{\rm b}$ decays are analyzed. The analysis expands earlier LHC results to lower masses. Upper limits at 95\% CL are presented in Figure~\ref{fig:BSM}(left).

\vspace{0.1cm}
\textbf{An experimental overview of EFT} was presented~\cite{Owen}. If no direct production of new resonances has been observed at the LHC, new physics at higher energy will still impact LHC measurements. This can be parameterised with an EFT approach. The challenge here is the number and the choice of the operators to consider. The aim is to perform global fits of many operators with many input physics measurements. The latest Higgs combine fit by ATLAS considers a total of 19 EFT parameters. 
EFT limits were also produced from a diverse range of analyses by ATLAS (${\rm t\bar{t}}$Z+${\rm t\bar{t}}\gamma$, WWjj, hh$\rightarrow {\rm b\bar{b}}\gamma\gamma$) and CMS ($\gamma\gamma\rightarrow\tau\tau$, ${\rm t\bar{t}}$+leptons).
 
\vspace{0.1cm}
\textbf{The concept of anomaly detection} proposes a new idea of model-agnostic searches at LHC that make use of deep learning tools to learn directly from data how the SM behaves~\cite{Ngadiuba}. This enables to eliminate signal priors and to search for anything anomalous with respect to the SM. 
A collection of complementary anomaly detection methods, based on unsupervised, weakly-supervised and semi-supervised algorithms, are used in order to maximise the sensitivity to unknown new physics signatures.
The methods were first applied to dijet and mulijet final states, and then extended to lepton/photon + jets. 
As an example, CMS tested different approaches as variational autoenconders (VAE), Cathode and Tag N'Train (TNT) and quasi anomalous knowledge (Quak) to perform a model-agnostic search for di-jet resonances with anomalous jet substructure. Different signals have been injected to test the methods and the results are presented in Figure~\ref{fig:BSM}(right).
For the first time at colliders, studies were also performed by CMS of incorporating anomaly detection early at the trigger level. CMS has developed two anomaly detection autoencoders for the first level (L1) trigger at the FPGA level. Firmware and stability tests were performed.

%
\section{Flavor physics} \label{sec:flavor}

Three half-day sessions were devoted at the conference to SM precision measurements and to BSM searches in the heavy flavour sector, covering a large number of analyses from the LHCb and Belle/BelleII collaborations. Results on flavour physics from CMS/ATLAS, NA62, BES III and KoTo were also discussed.  A selection of the latest results are presented in this section.

\vspace{0.1cm}
\textbf{The LHCb  and Belle/BelleII experiments} are dedicated to study flavour physics with great precision from the decays of B mesons. The integrated luminosity collected by LHCb during run 1 and run 2 is 3 and 6 fb$^{-1}$, respectively. Differently to the ATLAS and CMS upgrade program, the upgrade I of the LHCb detector took place before the run 3, and the detector upgrade II of LHCb is expected before the LHC run 5. 
The Belle detector, located at the collision point of the asymmetric-energy electron-positron collider KEKB, recorded collision data from 1999 to 2010 corresponding to an integrated luminosity of about one ab$^{-1}$. The BelleII experiment located at SuperKEKB recorded the run 1 dataset from 2019 to 2022, with an integrated luminosity of 0.42 ab$^{-1}$. The BelleII run 2 is presently ongoing. The target dataset luminosity for BelleII is 50 ab$^{-1}$. The main part of the Belle/BelleII data are collected at the centre of mass energy equals to the mass of the $ \Upsilon$(4S) resonance which decays to pairs of B mesons.

\vspace{0.1cm}
\textbf{Study of ${\rm b} \rightarrow {\rm s} \ell \ell$ transition} is particularly interesting as it has an extremely small amplitude in the SM, and is then an excellent place to search for new physics. Several analyses are performed as lepton flavour universality tests, measurement of differential branching ratios (d$\Gamma$/d$q^2$), or angular analysis ($P_5$', $C_9$, $A_{\rm FB}$ ...). As an example, the fit for the $C_9$ parameter, assuming SM values for other Wilson coefficients, presents a few $\sigma$ discrepancy with the SM prediction. 
However a difficult part of the SM computation comes from the long-distance term, in particular the charm loop correction which is difficult to calculate. 
LHCb presented a first unbinned amplitude analysis of B$^0 \rightarrow$ K$^{*0}$ (K$^+ \pi^-$) + $\mu^+ \mu^-$  with the run 1 and the 2016 dataset (4.7 fb$^{-1}$)~\cite{Smith} with the goal to try to measure from data the long-distance contribution with and without theory constraint. It is a model dependent analysis, that maximises sensitivity to non-local effects. The conclusion of the analysis is that even with the freedom of a non-local component, data prefer a shift in $C_9$ from the SM. 

\begin{figure}[t!]
\begin{minipage}{0.48\linewidth}
\centerline{\includegraphics[width=0.99\linewidth]{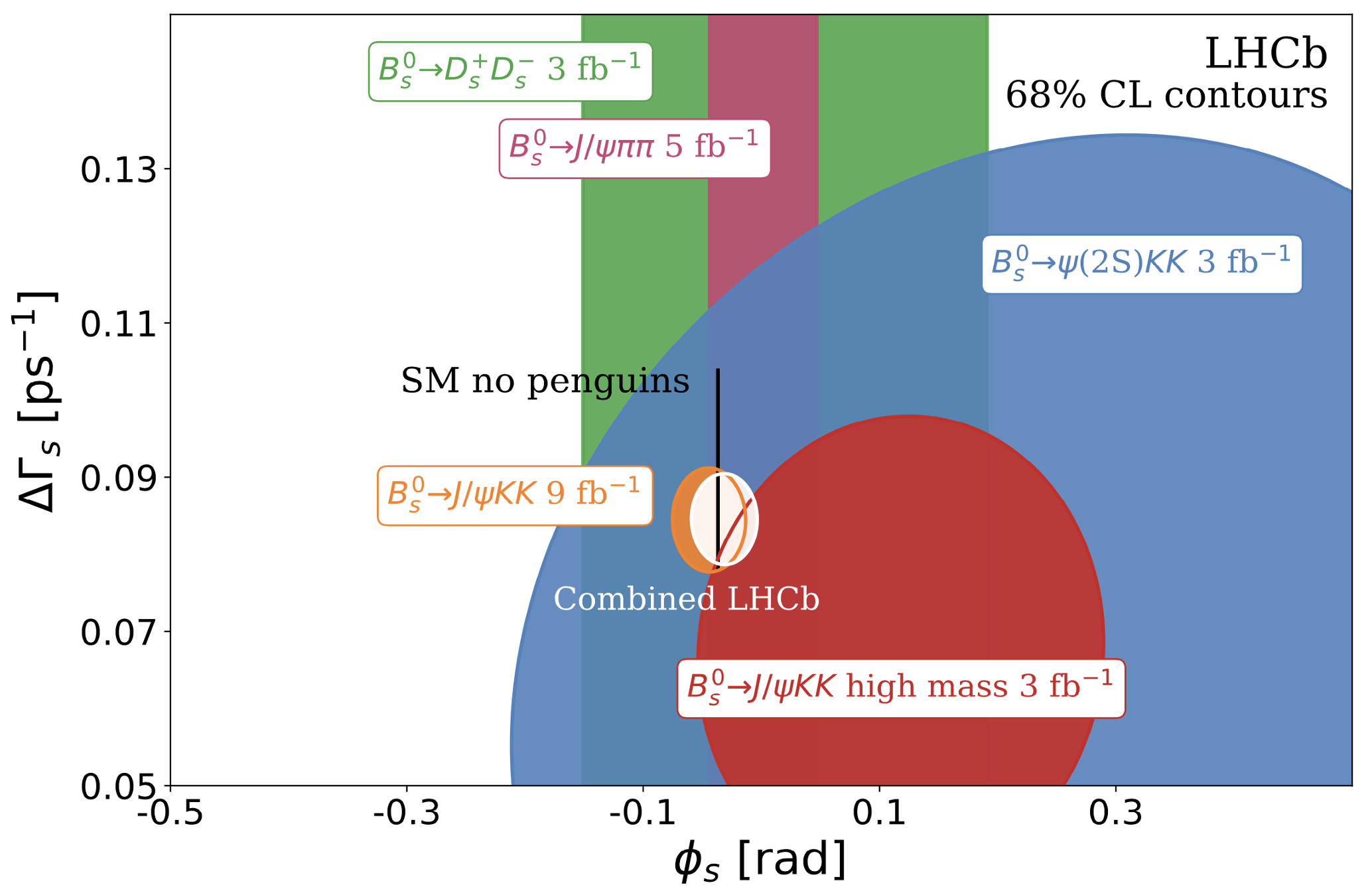}}
\end{minipage}
\hfill
\begin{minipage}{0.51\linewidth}
\centerline{\includegraphics[angle=0,width=0.9\linewidth]{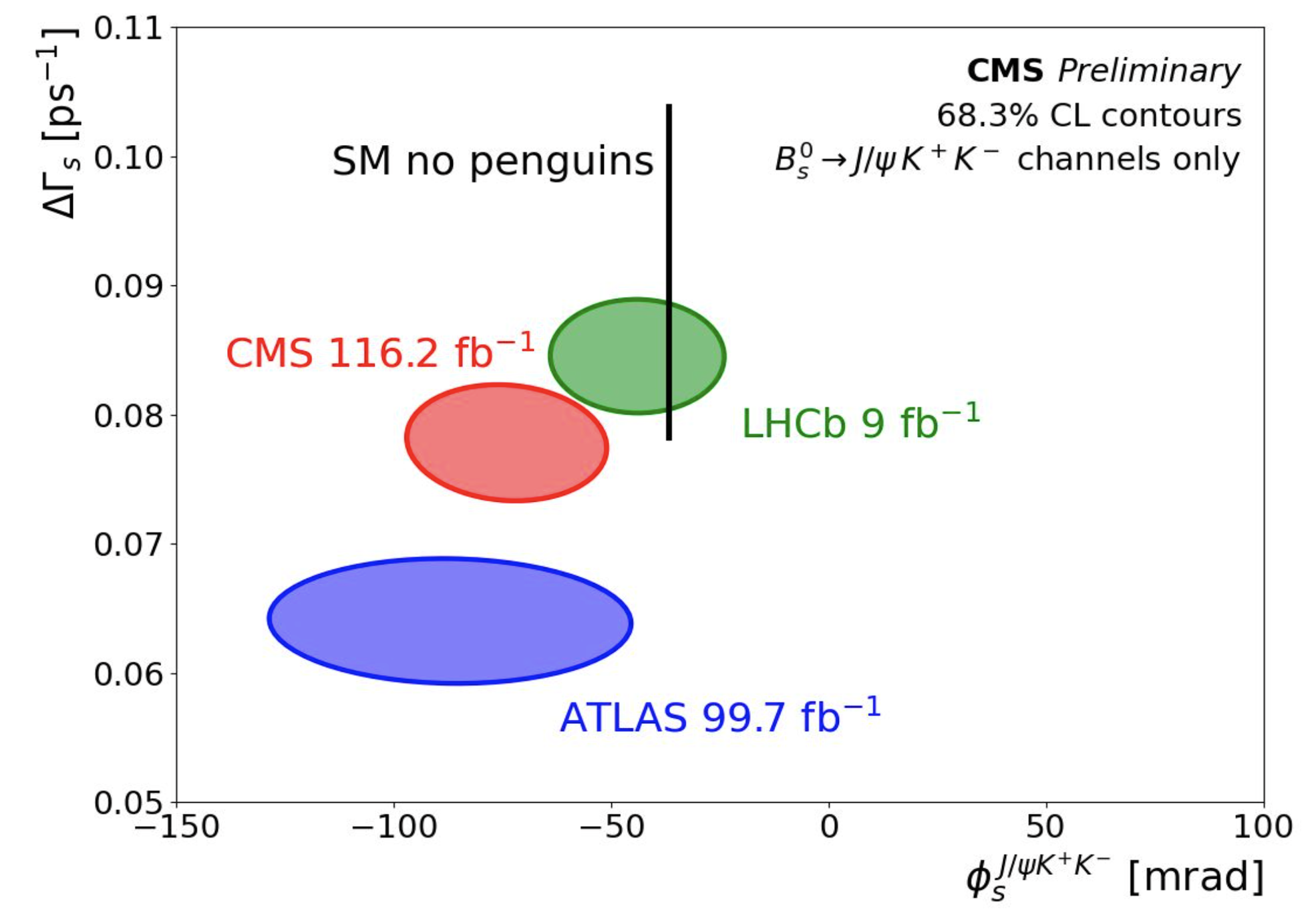}}
\end{minipage}
\caption[]{Measurements of the $\phi_{\rm s}$ angle by LHCb together with its combination (left) and comparison with the results from CMS and ATLAS (right).}
\label{fig:Flavour_LHCb}
\end{figure}

\vspace{0.1cm}
\textbf{Results on CP violation (CPV) in beauty and charm} were reported.
LHCb presented updated results on the measurement of the CKM angle $\gamma$, and on time dependent CPV in charm and beauty mesons. For the latter, the time dependent CPV decays of B$^0$, in the benchmark decay channels of J/$\psi {\rm K}^0_{\rm s}$ and $\psi$(2S)K$^0_{\rm s}$ were presented, leading to the measurement of the unitarity triangle $\beta$ angle~\cite{Williams}. 
The corresponding challenging analysis in the case of B$^0_{\rm s}$ was also performed in the golden decay channel J/$\psi$ K$^+$K$^-$, leading to the measurement of the $\beta_{\rm s}$ angle. 
In the SM, the weak phase $\phi_{\rm s} \approx -2 \beta_{\rm s}$,  and $\beta_{\rm s}$ is determined by the CKM global fit to be $\beta_{\rm s}= -37 \pm 1$ mrad. New physics can change the value of $\phi_{\rm s}$ up to about 100$\%$ via new particles contributing to the flavour oscillations. 
The LHCb final combination results, $\phi_{\rm s} = -31 \pm 18$ mrad, is shown in Figure~\ref{fig:Flavour_LHCb}(left) as the white circle, it is the most precise measurement to date, still statistically limited.

\vspace{0.1cm}
\textbf{CMS Wildcard -} The CMS collaboration also presented results on CPV measurement in B$^0_{\rm s}$ mesons in the same golden decay channel B$^0_{\rm s} \rightarrow$  J/$\psi \phi \rightarrow  \mu^+ \mu^-$ K$^+$K$^-$,
using the data collected in 2017-18 (corresponding to 98 fb$^{-1}$)~\cite{Bragagnolo}.  
The CMS analysis is a time, flavour and angular dependent measurement, using a new flavour tagging framework (based on 4 DNN-based algorithms). The result $\phi_{\rm s} = -74 \pm 23$ mrad provides evidence of CPV in  B$^0_{\rm s} \rightarrow$  J/$\psi$ K$^+$K$^-$ at 3$\sigma$ level.
The CMS result is presented in Figure~\ref{fig:Flavour_LHCb}(right).

\begin{figure}[t!]
\begin{minipage}{0.47\linewidth}
\centerline{\includegraphics[width=0.99\linewidth]{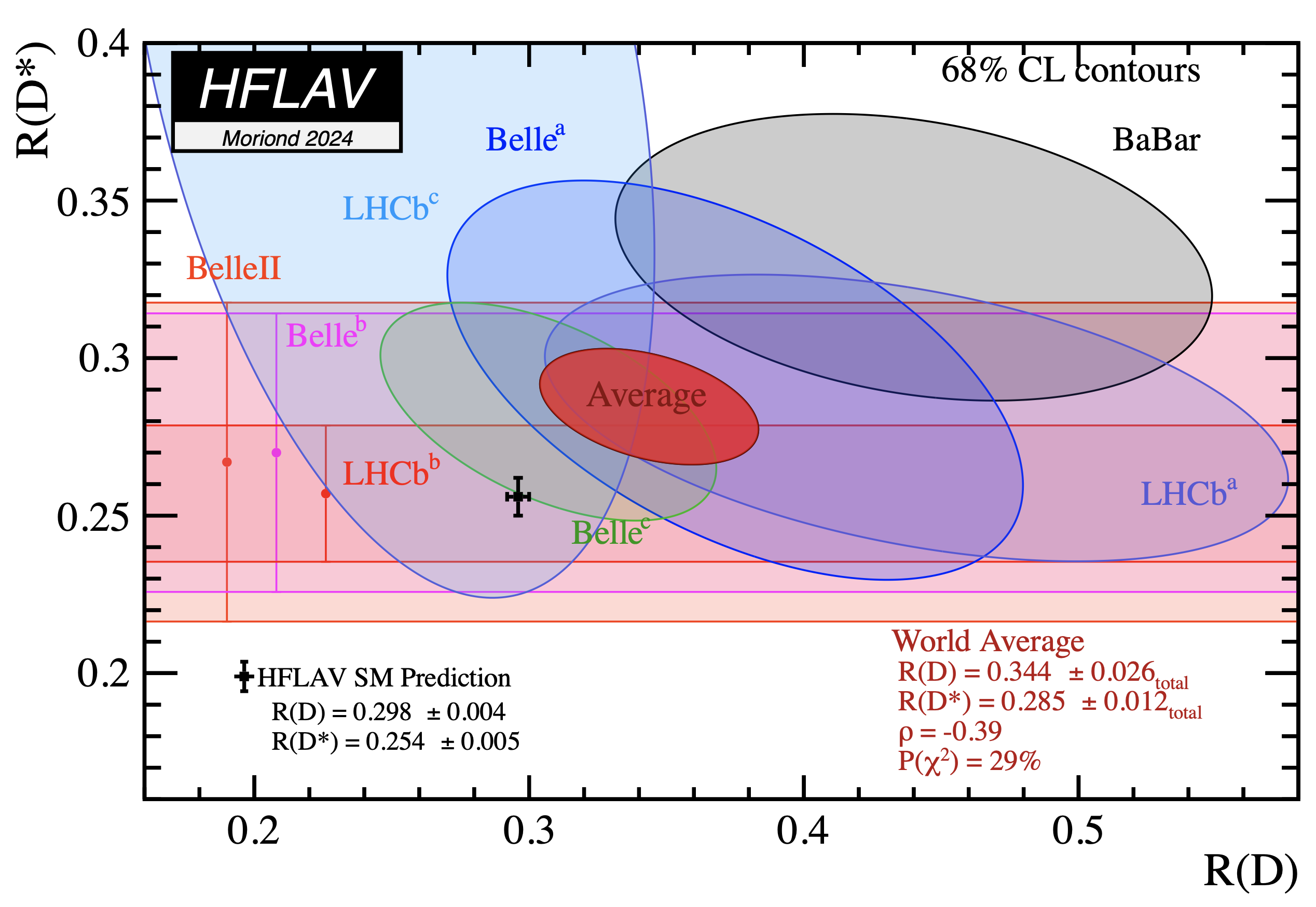}}
\end{minipage}
\hfill
\begin{minipage}{0.45\linewidth}
\centerline{\includegraphics[angle=0,width=0.999\linewidth]{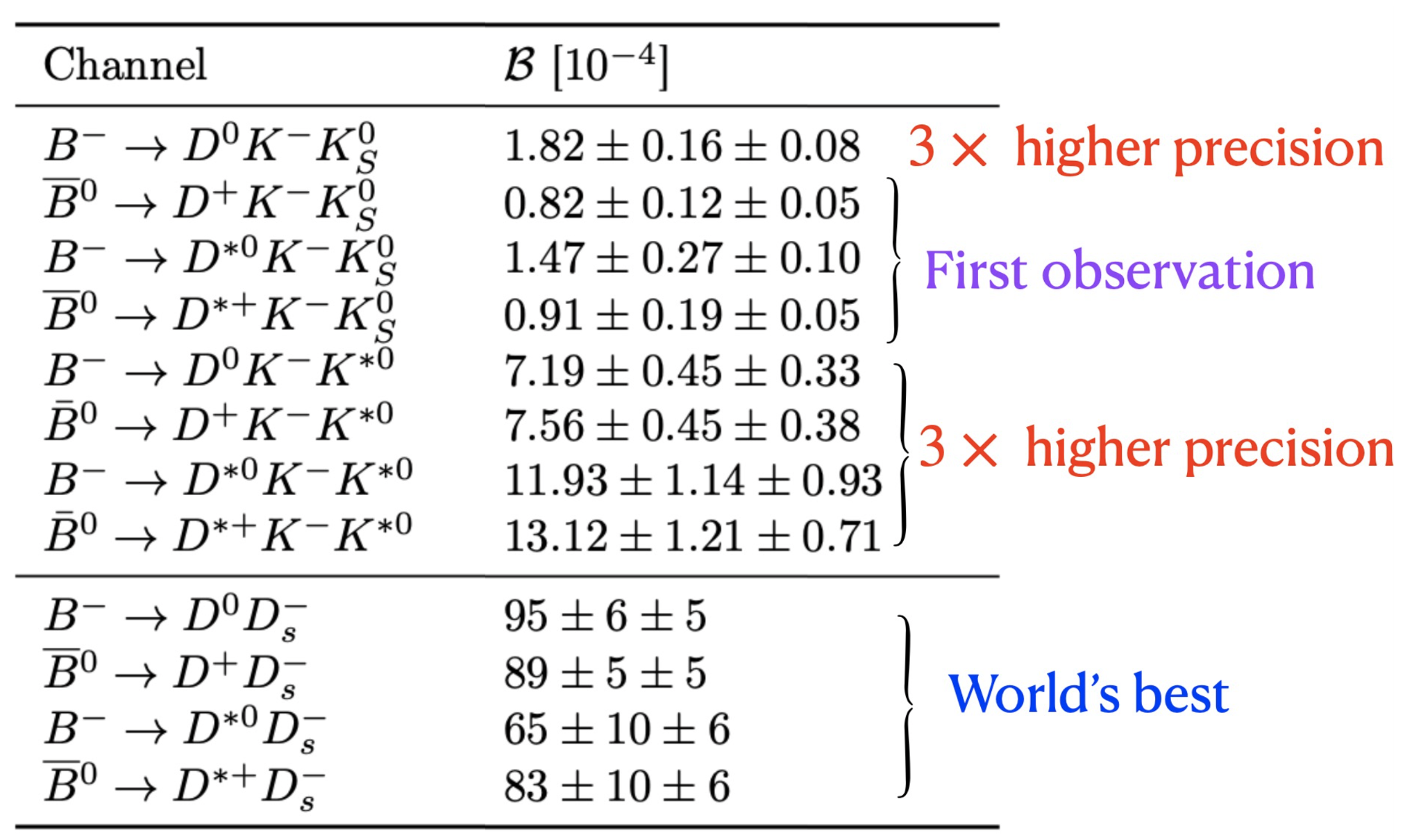}}
\end{minipage}
\caption[]{Left: Recent results from LHCb and BelleII for the measurements of R(D$^*$) and R(D). The new world average is represented by the red area and the SM expectation by the black cross; Right: Belle/BelleII latest results for hadronic B decay branching ratio measurements.}
\label{fig:Flavour_RD}
\end{figure}

\vspace{0.1cm}
\textbf{Study of ${\rm b} \rightarrow {\rm c} \ell \nu$ transition} was reported by LHCb~\cite{Pardinas} and Belle/BelleII~\cite{Cao}. New results were detailed on the ratio $R$ defined as  
$R($H$_ {\tau / \ell})=$BR(B$\rightarrow {\rm H} \tau \nu)$/BR(B$\rightarrow {\rm H}\ell \nu)$, where H = D or D* and $\ell= $e$, \mu$. These ratio measurements provide stringent tests on lepton flavour universality. In the ratio measurement, the normalisation to ($|V_{\rm cb}|$) cancels, as well as part of the theoretical and experimental uncertainties.
The LHCb results obtained using the 2015-16 data sample (2 fb$^{-1}$) are presented in Figure~\ref{fig:Flavour_RD}(left) in the (R$_{\rm D}$, R$_{\rm D^*}$) plane as the light blue area. The BelleII analysis (using the dataset of 189 fb$^{-1}$) for R$_{\rm D^*}$ is also shown. It is interesting to notice that the new world average is in tension with the SM at the level of 3.2$\sigma$.

\vspace{0.1cm}  
\textbf{Hadronic B decay measurements}  at Belle/BelleII are an active domain~\cite{Rout}. Several new measurements were reported, with a first observation for three channels and with improved precision for the others, as it is shown in Figure~\ref{fig:Flavour_RD}(right). 

%
%

\vspace{0.1cm}
\textbf{Latest results from BESIII - } Located at the BEPCII complex (beam energy from 1 to 2.5 GeV), the BESIII detector is operating since 2008. It is collecting the largest charmed hadron data samples collected near the mass threshold of charmed hadron pairs. Several new charm hadron BR decay measurements were presented at the conference~\cite{Liu}. As an example, the new BESIII measurement of  BR(D$^+_{\rm s}  \rightarrow \mu^+  \nu_\mu$) = $(5.29 \pm 0.11 \pm 0.09)$ $10^{-3}$ is the most precise to date. Using this measurement together with input from lattice QCD calculation, the CKM matrix element $|V_{\rm cs}|$ is determined with a precision of 1.4\%. When combined with the tau decay channel analysis, the precision on $|V_{\rm cs}|$ value improves to 1.0\%. Lepton flavour universality tests have also been performed in leponic and semi-leptonic decays of charm mesons. No violation was observed at the 1.5\% precision level.

\section{Neutrino physics} \label{sec:neutrinos}

Neutrino physics is a very active research domain, with a series of running and planed experiments. Indeed, if we have measured fairly well the PMNS mixing matrix elements ($\theta_{12},\theta_{23},\theta_{13}$) and the square of the neutrino mass differences ($\Delta m^2_{21}$ and $\Delta m^2_{32}$) from the neutrino oscillation experiments, some important open questions stay. We still need to measure the neutrino mass ordering (NMO), the CP violation phase $\delta_{CP}$ and the octant degeneracy, as well as the neutrino masses. Another key question is the nature of the neutrino, if it is a Dirac or a Majorana particle. If the neutrinos are Majorana particles, they are their own antiparticles. This has implications for the understanding of fundamental symmetries in physics as the matter-antimatter asymmetry in the universe. During the conference, results were presented from two long baseline running experiments NOvA and T2K dedicated to study neutrino oscillations, and from the Super-Kamiokande experiment detecting atmospheric neutrinos. These results are summarised below, with an update on the neutrinoless double-beta decay status from CUORE and Legend, as well are the first results from the FASER experiment. 

\vspace{0.1cm}
\textbf{The NOvA and T2K experiments} are two neutrino long baseline experiments with an oscillation distance of 810 km (from Fermilab to Minnesota in USA) and 295 km (from JPARC to Super-Kamiokande in Japan) respectively, and a beam peak energy of 2 GeV for NOvA and 0.6 GeV for T2K. For both experiments, the muon neutrino disappearance and the electron neutrino appearance are the signature of neutrino oscillations. The NOvA experiment started to take data in 2014. T2K took data from 2010 to 2021. After a beam power upgrade and an acceptance improvement of the near detector (ND80), the second phase of the T2K data taking started in February 2024.

\begin{figure}[t!]
\begin{minipage}{1.0\linewidth}
\centerline{\includegraphics[width=0.99\linewidth]{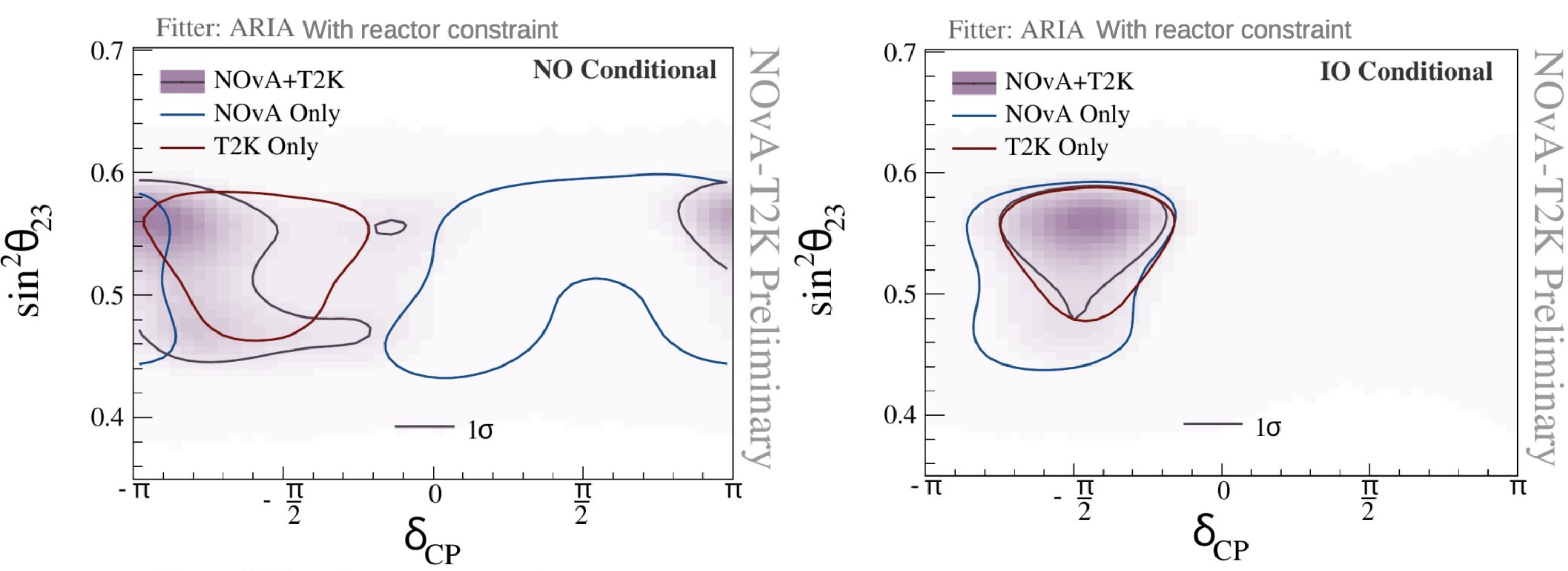}}
\end{minipage}
\caption[]{Neutrino oscillation results from NOvA and T2K together with the results from the combined fit, for the normal (left) and the inverted (right) neutrino mass ordering hypotheses.}
\label{fig:Neutrino_comb}
\end{figure}

\begin{figure}[b!]
\begin{minipage}{0.55\linewidth}
\centerline{\includegraphics[width=0.99\linewidth]{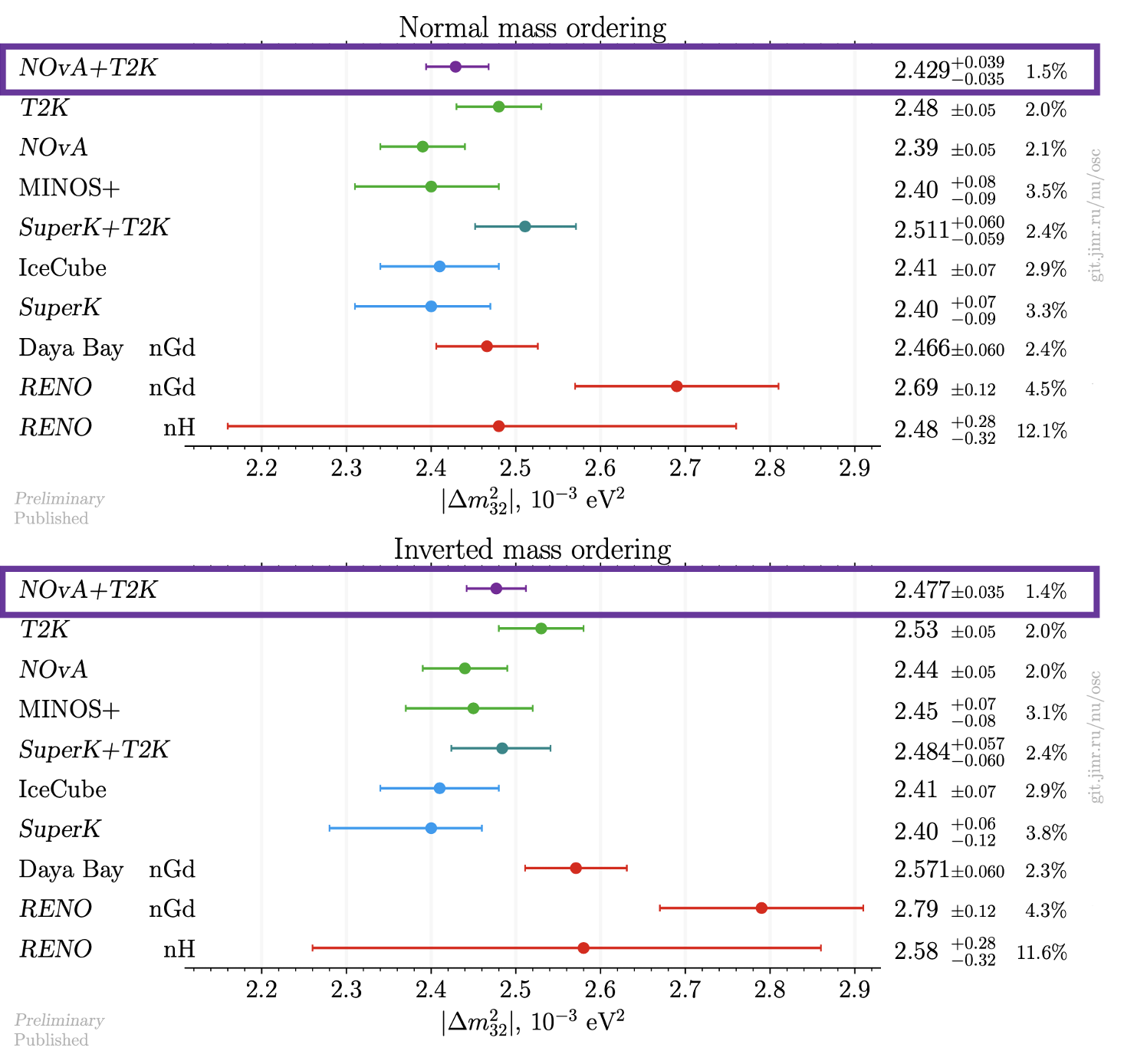}}
\end{minipage}
\hfill
\begin{minipage}{0.43\linewidth}
\centerline{\includegraphics[angle=0,width=0.9\linewidth]{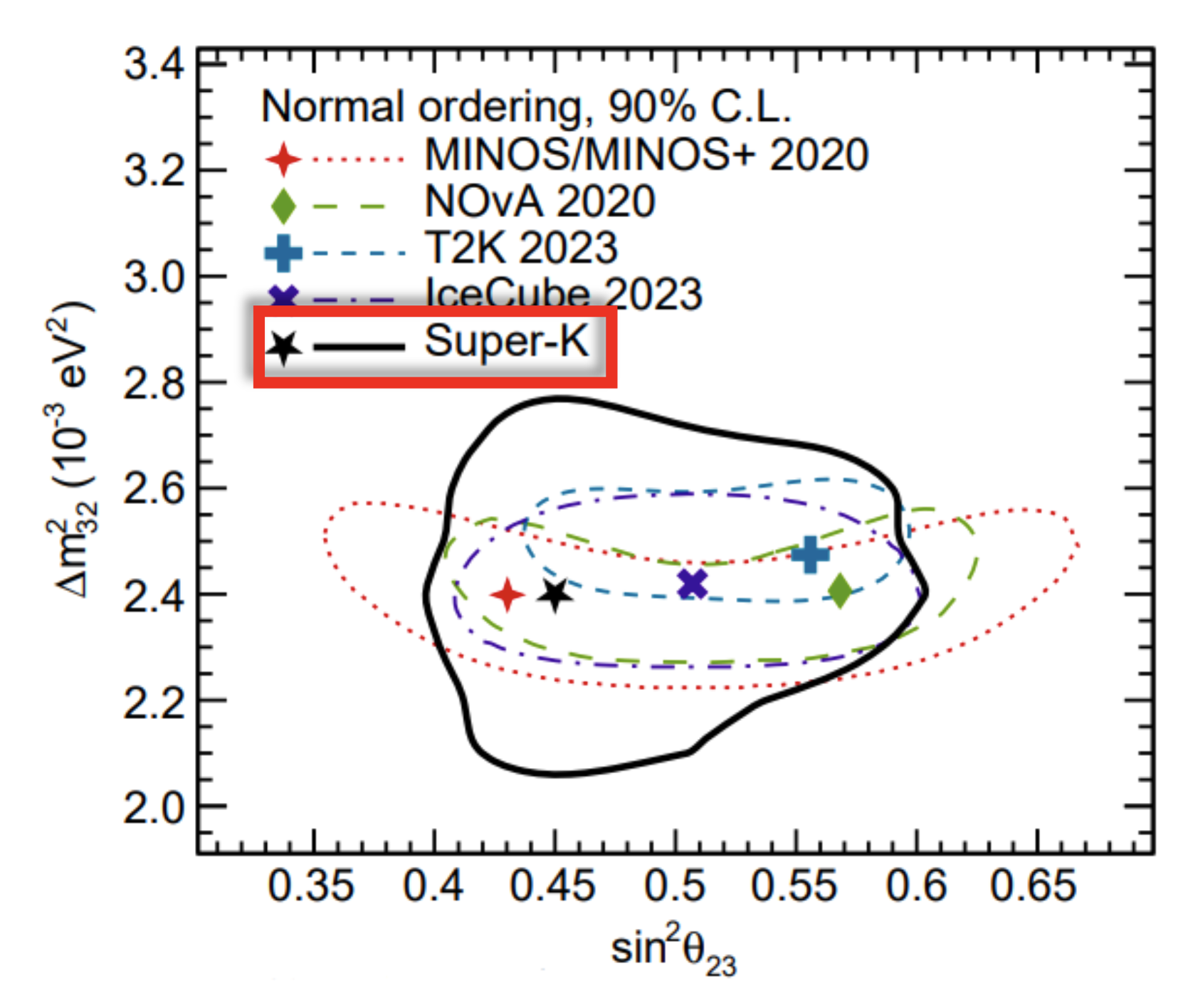}}
\end{minipage}
\caption[]{Latest measurements of the oscillation parameter $\Delta m^2_{32}$ from NOvA and T2K, together with the value obtained from the combined fit (left); Results on $\Delta m^2_{32}$ and sin$^2\theta_{23}$ measurements from the Super-Kamiokande experiment (right).}
\label{fig:Neutrino_comb_superK}
\end{figure}

\vspace{0.1cm}
\textbf{Combining NOvA and T2K} data are interesting as their complementarity can resolve degeneracies, the T2K measurement isolates the impact of CP violation whereas NOvA has significant mass ordering sensitivity. Lots of efforts went into the combination procedure as the experiments have different analysis approaches driven by contrasting experimental designs~\cite{Sanchez}. The joint fit uses data collected by each experiment up to year 2020. The individual NoVA and T2K results, together with their combination, are presented in Figure~\ref{fig:Neutrino_comb}. If the experiments individually have a preference for the normal ordering (NO), the joint fit has a weak preference for the inverted ordering (IO). So the mass ordering remains inconclusive. The NO allows for a broad range of $\delta_{CP}$ values, while in the case of IO, CP conserving values lie outside the 3$\sigma$ interval. 
The joint analysis gives also a strong constraint on the absolute value of $\Delta m^2_{32}$ as shown in Figure~\ref{fig:Neutrino_comb_superK}(left).

\vspace{0.1cm}
\textbf{The Super-Kamiokande (SK) experiment}, a large water detector surrounded by about 11000 PMTs in the inner part and with an outer muon veto detector, is running since 1996, detecting atmospheric neutrinos. The water volume was gadolinium-doped since 2020 for easier neutron capture identification, allowing physicists to separate neutrino from antineutrino interactions. The SK presented competitive measurements compared with other experiments, in particular for the angle $\theta_{23}$, as shown in Figure~\ref{fig:Neutrino_comb_superK}(right)~\cite{Santos}.

\vspace{0.1cm}
\textbf{Combining T2K and SK} data are interesting as the SK helps in breaking T2K degeneracy between the NMO and the CP phase~\cite{Lichfield}. The combined results show that CP conservation is disfavoured by about 2$\sigma$.

\vspace{0.1cm}
\textbf{Searching for neutrinoless double-beta (0$\nu\beta\beta$) decay} is performed by several experiments to shed light on the nature of neutrinos. The reaction to be searched for is (A,Z) $\rightarrow$ (A, Z+2) + 2e$^-$. Results from the CUORE~\cite{Nutini} and Legend~\cite{Barton} experiments were presented at the conference. CUORE is a cryogenic experiment at tonne-scale, located at Gran Sasso in Italy, utilising oxide of tellurium (TeO$_2$) thermal detectors operated at about 10 mK.
The challenges of the experiment are to keep low temperature and low vibrations over time for about 1000 detectors, as well as low backgrounds. 
The limit obtained for the half-live time of $^{130}$Te, based on data taken from 2017-2023 (corresponding to an exposure of 2039 kg yr), is T$_ {0\nu}^{1/2} > 3.8$ x $10^{25}$ yr at 90\% CL, which is the most stringent limit for the  $^{130}$Te to date. The corresponding limit on the effective Majorana mass assuming a light Majorana neutrino-exchange is m$_{\beta\beta}<$ 70-240 meV.  The Legend experiment using enriched germanium detector $^{76}$Ge has a dual-phase program: Legend-200 (which is now operational with 142/200 kg of detector mass) and Legend-1000 (planned). Analysis techniques from the previous GERDA and Majorana  demonstrator experiments are being modified and applied to Legend data, with new analysis techniques in active development. The ultimate goal is to reach sensitivity for a half-life time of this nucleus beyond 10$^{28}$ years, corresponding to a neutrino effective mass measurement of about 18 meV.

\begin{figure}[h!]
\begin{minipage}{0.33\linewidth}
\centerline{\includegraphics[width=0.99\linewidth]{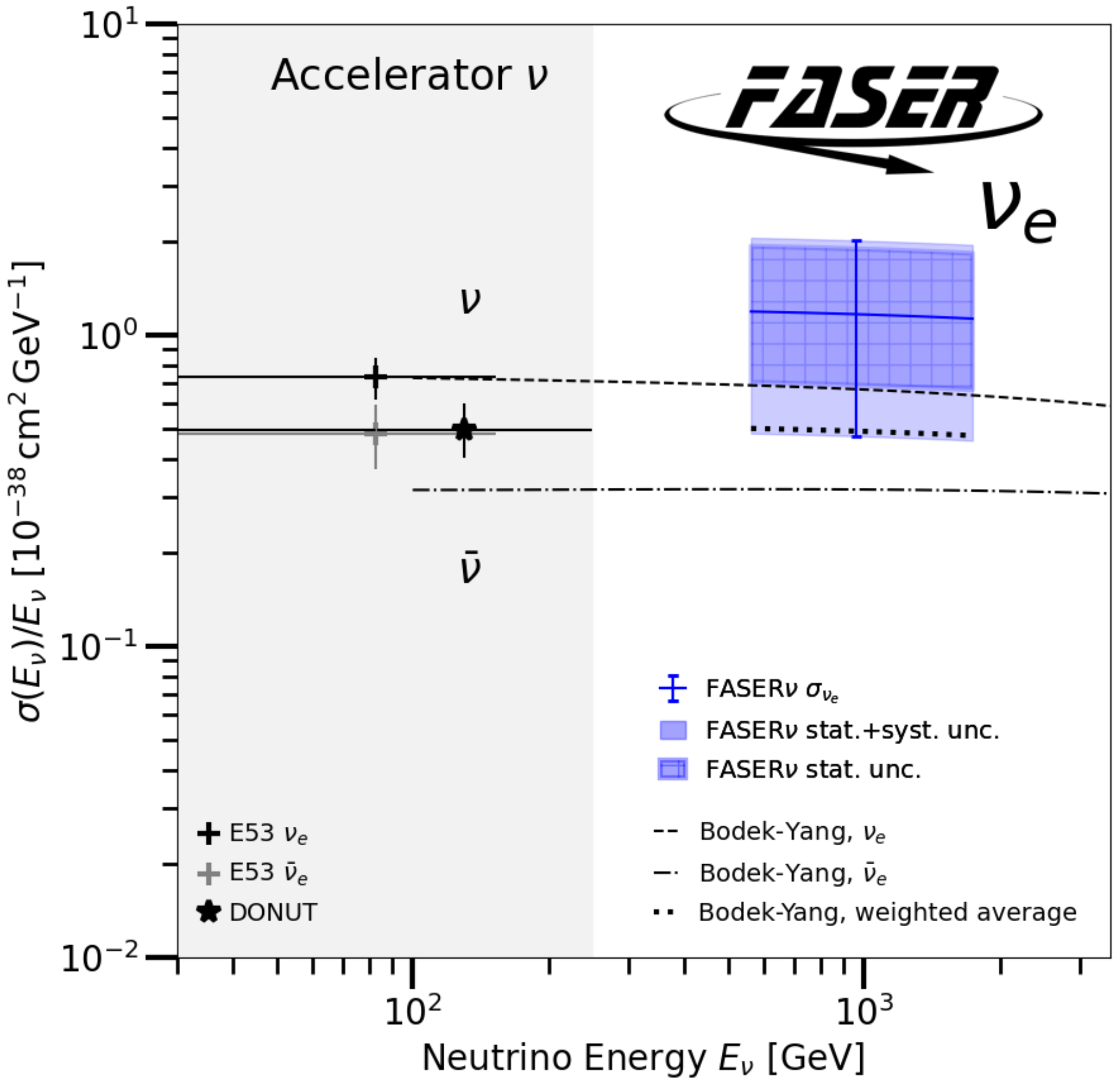}}
\end{minipage}
\hfill
\begin{minipage}{0.65\linewidth}
\centerline{\includegraphics[angle=0,width=0.9\linewidth]{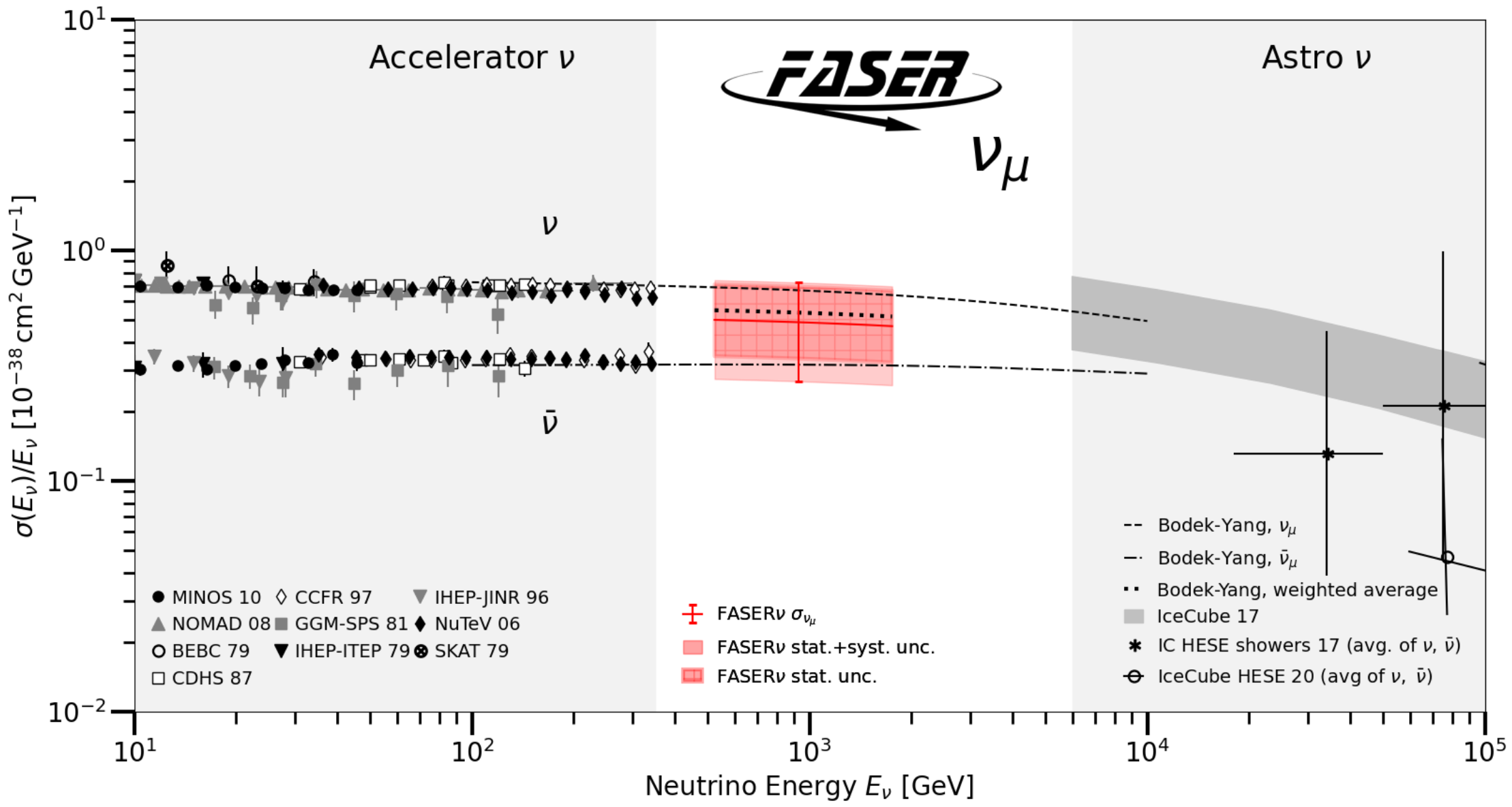}}
\end{minipage}
\caption[]{The Faser $\nu_e$ (left)  and $\nu_\mu$(right) cross section measurements as function of the neutrino energy. }
\label{fig:Neutrino_faser}
\end{figure}

\vspace{0.1cm}
\textbf{The Faser experiment} is a forward search experiment at the LHC, located 480 m downstream of the ATLAS interaction point. Its goal is to measure SM neutrino interaction cross sections at unexplored TeV energies, as well as to search for long-lived BSM particles (e.g. axion-like particles ALPs). The experiment is taking data in the LHC run 3, with about 70 fb$^{-1}$ collected to date. New results were presented~\cite{Ariga} on neutrino ($\nu_e$ and $\nu_\mu$) interaction cross sections using only 2\% of the run 3 collected data, as shown in Figure~\ref{fig:Neutrino_faser}. This represents the first detection of $\nu_e$ at the LHC. Results on limits on ALPs were also shown for a luminosity of 57.7 fb$^{-1}$, excluding uncovered parameter space (in the coupling and mass plane) significantly. The Faser data taking during run 4 has been approved.

\vspace{0.1cm}
Other updates on neutrino sector were given during the conference. The present neutrino mass limit of 0.8 eV from the \textbf{Katrin experiment} was reminded and the future release of the neutrino mass limit with 0.5 eV sensitivity expected for mid-2024~\cite{Marsteller} was presented, together with the R\&D for the Katrin$^{++}$ project to reach the inverted ordering mass scale. The Katrin collaboration also presented a dedicated sterile neutrino search with Tristan (tritium sterile anti-neutrino). An interesting (small) deficit of events was observed by the \textbf{IceCube experiment}~\cite{Hardin} in the muon antineutrino survival probability for atmospheric neutrinos, that can be fitted with the addition of a fourth neutrino family (the $p$-value for the null hypothesis is 3.1\%). More data will be available in particular with the addition of the dense array of 7 strings to be deployed in early 2026 (the phase 1 upgrade).

%

\section{Dark matter searches and precision measurements} \label{sec:DM}

\vspace{0.1cm}
\textbf{Searches for dark matter (DM)} by the Lux-Zeplin experiment~\cite{Rischbieter} at the Sanford Underground Research Facility and by the PandaX experiment~\cite{Zhou} at the China Jinping Underground Laboratory were reported. Both experiments are designed for WIMP (weakly interacting massive particle) direct detection with a dual phase xenon time-projection chamber (TPC) detector, measuring both the scintillation and the ionization signal responses. The experiment's characteristics are the precise energy measurement, the 3D event position reconstruction and the discrimination between nuclear recoil and electron recoil signals. The Lux-Zeplin experiment presented the spin-independent WIMPS-nucleon cross section limit as a function of the WIMP mass, from data collected between December 2021 and May 2022, and corresponding to 5.5 $\pm$ 0.2 tonne of fiducial volume. PandaX is in its commissioning phase which started in November 2020. Both experiments showed new results beyond the WIMP searches, as residual weak electromagnetic properties or non-relativistic effective field theory.

\vspace{0.1cm}
\textbf{Updates on anomalous magnetic moment of the muon} defined as $a_\mu=(g_\mu-2)/2$ were also discussed~\cite{Lehner}. It is a very sensitive variable to new physics, as the quantum effects arise from virtual particle contributions from all known and potentially unknown particles. The long-standing discrepancy between the experimental measurements and the theory predictions has been scrutinised during the conference. The Fermilab Muon g-2 experiment is providing improved measurements, currently at a precision of 0.2 ppm. A lot of efforts are dedicated to the SM calculation, and more specifically on the hadronic vacuum polarisation contribution. New results on lattice QCD have been presented and when taken into account, the SM prediction for $a_\mu$ falls better in line with the experimental results. However these computations are complicated, and lattice QCD results from other groups are expected to be public soon. A discussion will then take place on the inclusion or not of these results in the official SM calculation.


\section{Conclusions} \label{sec:conc}
A rich amount of latest results obtained in the field of high energy physics was presented at the conference and triggered lively and interesting discussions among the 140 participants. The many new results presented show that both experimental and theoretical physicists continue to progress on all fronts. If the SM is still in very good shape, a few slight deviations observed in various sectors are scrutinised. Concerning the future of the large experiments, more and more data are being collected, already now with upgraded beam/detectors or will come in the short future, covering the various sectors of the field. Lets mention for example the currently running upgraded LHCb and BelleII experiments on B physics; the presently running ATLAS and CMS (run 3) and the coming high luminosity LHC phase to probe the SM, in particular the H boson sector and searches for new physics; the upgraded T2K experiment and the planed experiments in the neutrino sector as Hyper-Kamiokande, ORCA/JUNO, and Dune; the Lux-Zeplin and PandA experiments searching for direct signal of dark matter, and many others. This illustrates the bright future of the field and certainly important new results to be released in the coming year(s).
 
\section*{Acknowledgments}

The author is grateful to the speakers and participants to the conference for lively and inspiring discussions, to the Moriond committee and support staff for organising such an interesting program and beautiful conference. The author acknowledges the support of the Belgian Fund for Scientific Research, the F.R.S.-FNRS (Fond pour le Recherche Scientifique). 
 
%
%
%


\section*{References}

\begin{thebibliography}{99}
%
\bibitem{Devivie} J.-B. de Vivie, {\em H mass and width in ATLAS and CMS}, in this proceeding.
\bibitem{Trevisani} N. Trevisani, {\em Higgs coupling measurements in ATLAS and CMS}, in this proceeding.
\bibitem{Mohammadi} A. Mohammadi, {\em Differential Higgs cross sections in ATLAS and CMS}, in this proceeding.
\bibitem{Deramo} L. d'Eramo, {\em Di-Higgs searches in ATLAS and CMS}, in this proceeding.
\bibitem{Uttley} G. Uttley, {\em Search for BSM Higgs in ATLAS and CMS}, in this proceeding.
%
\bibitem{Long} K. Long, {\em Electroweak results from ATLAS and CMS}, in this proceeding.
\bibitem{Merli} A. Merli, {\em Electroweak results of LHCb}, in this proceeding.
\bibitem{Khukhunaishvili} A. Khukhunaishvili, {\em CMS Wildcard: $\sin^2\theta^\ell_ {\rm eff}$ and FB asymmetry}, in this proceeding.
\bibitem{Knue} A. Knue, {\em ATLAS Wildcard: Test of LFU in W boson decay}, in this proceeding.
\bibitem{Caillol} C. Caillol, {\em Electroweak results in two photon collisions}, in this proceeding.
%
\bibitem{Wuchterl} S. Wuchterl, {\em Top quark properties from ATLAS and CMS}, in this proceeding.
\bibitem{Boumediene} D. Boumediene, {\em Associated top production in ATLAS and CMS}, in this proceeding.
\bibitem{Bragagnolo} A. Bragagnolo, {\em CMS wildcard - Probing fundamental properties of nature in heavy quark physics in CMS}, in this proceeding.
%
\bibitem{Grancagnolo} S. Grancagnolo, {\em Search for VLQ, HNL, LLP in ATLAS and CMS}, in this proceeding.
\bibitem{Santanasasio} F. Santanastasio, {\em Other exotics searches by CMS}, in this proceeding.
\bibitem{Kay} E. Kay, {\em Other exotics searches by ATLAS}, in this proceeding.
\bibitem{Clement} C. Cl\'ement, {\em ATLAS search for new spin-0 resonances in X$\rightarrow$SH$\rightarrow$b$\bar{b}\gamma\gamma$}, in this proceeding.
\bibitem{Owen} M. Owen, {\em Experimental overview of EFT in ATLAS and CMS}, in this proceeding.
\bibitem{Ngadiuba} J. Ngadiuba, {\em Model independent searches at the LHC}, in this proceeding.
%
\bibitem{Smith} M. Smith, {\em Results from LHCb on b$\rightarrow$s$\ell\ell$ }, in this proceeding.
\bibitem{Williams} M. Williams, {\em LHCb results on CKM and CPV in beauty and charm decays}, in this proceeding.
\bibitem{Pardinas} J. Pardinas, {\em Results from LHCb on b$\rightarrow$c$\ell\nu$}, in this proceeding.
\bibitem{Cao} L. Cao, {\em Results on Belle II on LFU in b$\rightarrow$c$\ell\nu$ }, in this proceeding.
\bibitem{Rout} N. Rout, {\em Hadronic B decays at Belle and Belle II}, in this proceeding.
\bibitem{Casarosa} G. Casarosa, {\em Charm physics at Belle and Belle II}, in this proceeding.
\bibitem{Goldenzweig} P. Goldenzweig, {\em Radiative and electroweak penguin results  from Belle and Belle II}, in this proceeding.
\bibitem{Kovanda} O. Kovanda, {\em B physics in ATLAS and CMS}, in this proceeding.
\bibitem{Liu} P. Liu, {\em Latest results from BESIII: charm hadron BR decay measurements}, in this proceeding.
%
\bibitem{Santos} A. Santos, {\em Latest results from Super-Kamiokande}, in this proceeding.
\bibitem{Lichfield} P. Lichfield, {\em Lastest results from T2K experiment}, in this proceeding.
\bibitem{Sanchez} M. Sanchez, {\em NOvA and T2K joint results}, in this proceeding.
\bibitem{Nutini} I. Nutini, {\em Searching for 0v$\beta\beta$ in the CUORE experiment}, in this proceeding.
\bibitem{Barton} C. Barton, {\em The Legend experiment}, in this proceeding.
\bibitem{Ariga} A. Ariga, {\em The Faser experiment}, in this proceeding.
\bibitem{Marsteller} A. Marsteller, {\em The Katrin experiment}, in this proceeding.
\bibitem{Hardin} J. Hardin, {\em Search for new physics in IceCube}, in this proceeding.
%
\bibitem{Rischbieter} G. Rischbieter, {\em Results form Lux-Zeplin}, in this proceeding.
\bibitem{Zhou} N. Zhou, {\em Results from PandaX}, in this proceeding.
\bibitem{Lehner} C. Lehner, {\em Status of muon g-2}, in this proceeding.
%
\end{thebibliography}
\end{document}